\documentclass{aa}  

\usepackage{graphicx}
\usepackage{txfonts}
\usepackage{enumitem}

\begin{document}

\title{Time evolution of gaps in stellar streams in axisymmetric St\"ackel potentials}
  \titlerunning{Evolution of gaps in St\"ackel potentials}
   
   \author{Helmer H. Koppelman\inst{1,2}
          \and Amina Helmi\inst{1}
          }

        \institute{
        Kapteyn Astronomical Institute, University of Groningen, Landleven 12, 9747 AD Groningen, The Netherlands \and 
        School of Natural Sciences, Institute for Advanced Study, Princeton, NJ 08540, USA\\
              \email{koppelman@ias.edu}
}
   \date{}

  \abstract
  {When a subhalo interacts with a cold stellar stream it perturbs its
  otherwise nearly smooth distribution of stars, and this leads to the
  creation of a gap. The properties of such gaps depend on the parameters
  of the interaction. Their characterisation could thus lead to the
  determination of the mass spectrum of the perturbers and possibly
  reveal the existence of dark subhalos orbiting the Milky Way.}
  {Our goal is to construct a fully analytical model of the formation and evolution of gaps embedded in streams orbiting in a realistic Milky Way potential.}
  {To this end, we extend the model of Helmi~\&~Koppelman~(2016) for spherical
  potentials, and predict the properties of gaps in streams evolving in
  axisymmetric St\"ackel potentials. We make use of
  action-angles and their simple behaviour to calculate the divergence
  of initially nearby orbits slightly perturbed by the interaction
  with a subhalo.}
  {Our model, corroborated by N-body experiments, predicts that the size
  of a gap grows linearly with time. We obtain analytical expressions
  for the dependencies of the growth rate on the orbit of the stream,
  the properties of the subhalo (mass, scale-radius), and the geometry
  of the encounter (relative velocity, impact parameter). We find that
  the density at the centre of the gap decreases with time as a
  power-law in the same way as the density of a stream. This results in the
  density-contrast between a pristine and a perturbed stream on the
  same orbit asymptotically reaching a constant value dependent only
  on the encounter parameters.}
  {We find that at a fixed age, smallish gaps are sensitive mostly to the mass of the subhalo, while gaps formed by subhalo flybys with a low relative velocity, or when the stream and subhalo move parallel, are degenerate to the encounter parameters.}


   \maketitle

\graphicspath{{./Figures/}}

\section{Introduction}
The widely accepted $\Lambda$CDM model is very successful in reproducing the large-scale structure of the Universe \cite[e.g.][]{Davis1985TheMatter}, but it faces some key problems on small scales \citep[e.g.][]{Bullock2017}. For example, on the scales of individual galaxies, we observe much less substructure than what is predicted by dark matter only cosmological simulations \citep{Klypin1999WHERE, Moore1999}. Such simulations show that substructure exists down to very small scales and can be found at all radii, although preferentially in larger numbers in the outskirts of galaxies' halos \citep[e.g.][]{Diemand2008ClumpsDistribution,Springel2008}. 

There exist several possibilities to solve this missing substructure conundrum. For example, adding baryonic physics to the simulations alleviates some of the problems, although mostly in the inner part of galaxies \citep[e.g.][]{DOnghia2010SubstructureDisk, Zhu2016BaryonicSatellites, Sawala2017}. Adjusting the properties of the dark matter particle (e.g. self-interacting dark matter, warm dark matter, or fuzzy dark matter) can help in suppressing the formation of the smallest substructures \citep[e.g.][]{Spergel2000ObservationalMatter,Hu2000FuzzyParticles,Bode2001HaloModels,Vogelsberger2016,Bozek2016ResonantUniverse,Hui2017UltralightMatter}. 
Another solution is to assume that the structures are present but in a dark form. Dark structures only reveal their presence through gravitational interaction, rendering them very difficult to detect. Results from gravitational lensing support the existence of dark structures at a level that is compatible with $\Lambda$CDM \citep{Dalal2002DIRECTSUBSTRUCTURE, Vegetti2010DetectionImaging, Vegetti2012GravitationalDistance, Ritondale2019Low-massCDM, Hsueh2020SharpQuasars}. 

Establishing whether there exists a population of subhalos with masses $<10^{8}~{\rm M}_\odot$ in and around the Galaxy is therefore of the utmost importance as it can lead to a better understanding of the nature of the dark matter particle. Clearly, the discrepancy between the predicted and observed small-scale structure could be hinting at a fundamental problem with our current cosmological paradigm. 

In this work, we will focus on a method to indirectly detect dark subhalos in our own Galaxy, through their possible interactions with cold stellar streams. Such streams are thin, almost one-dimensional elongated structures consisting of stars that originate from the tidal disruption of globular clusters or small dwarf galaxies. Because of their fragile nature, these streams are easily perturbed by gravitational interactions, making them promising probes of dark substructures \citep{Ibata2002UncoveringStreams, Johnston2002}. Occasional flyby's of dark subhalos can lead to the creation of a gap in an otherwise relatively smooth distribution of stars \citep{Yoon2011ClumpyStructures,Carlberg2013}. Unfortunately, finding streams is challenging because of their low surface brightness, let alone finding gaps in streams. However, recent deep photometric surveys have identified a few dozen of narrow streams \citep[e.g.][]{Belokurov2006, Bernard2016,Shipp2018a}. The analysis of {\it Gaia} DR2 \cite{GaiaCollaboration2016TheMission, GaiaCollaboration2018brown} has also yielded another dozen streams \citep{Malhan2018GhostlyCatalogue, Ibata2019TheGalaxy}. So far, only two of these streams have been claimed to contain gaps: GD-1 \citep{Grillmair2006DETECTIONSURVEY} and Palomar 5 (or Pal 5) \citep{Odenkirchen2001}, although several other streams show peculiarities \citep{Bonaca2019Jhelum, Shipp2019propermotions,Li2020}.

GD-1, is a promising stream to probe for gaps because of its length and coldness. It is known to contain several non-smooth features \citep{Carlberg2013GAPSSTREAM, DeBoer2018AWiggles, Price-Whelan2018OffStream}. The origin of these features, or gaps, is currently highly debated in the literature. For example, they could have been formed by an interaction with a massive (dark) object of $10^6 - 10^8 \, {\rm M}_\odot$ that might have once been part of the Sagittarius system \citep[][see also \citealt{Banik2019EvidenceStream}]{Bonaca2019TheHalo, Bonaca2020High-resolutionSagittarius}.
On the other hand, it has been argued that the presence and nature of a nearly periodic spatial distribution of gaps is an indication that these could be explained by internal dynamics without the need to recur to interactions with dark structures \mbox{\citep{Ibata2020DetectionSubhalos}}.

Pal 5's stream has been tentatively shown to host two gaps and several other features that would be consistent with being induced by subhalos in the range of $10^6 - 10^8 \, \mathrm{M_\odot}$ \citep{Erkal2017ATechnique, Bovy2017LinearSpectrum}. The inferred number of interactions appears to agree with the expected number predicted by CDM-only simulations \citep[e.g.][]{Sanderson2016}. Unfortunately, Pal 5's stream is not ideally suited to look for gaps due to dark structures because of its proximity to the Galactic Centre. The high baryon density in this region can lead to the formation of irregularities in the stream's profile, for example, due to interactions with the bar \citep{Pearson2017GapsRotation}, globular clusters and with other baryonic structures \citep{Banik2018EffectsStream}. Moreover, some of the gaps and features found in Pal 5's stream may be explained by survey incompleteness \citep{Thomas2016}. 

Since the expectation is that in the near future many gaps in many
different streams will be detected, it is imperative to develop an
in-depth understanding of the characteristics and evolution of these
gaps. With such an understanding we may be able to link the population of gaps to
an underlying population of dark substructures. For example, we need to
establish the relation between subhalos and gap sizes, the growth
rate of gaps and the dependence of their properties on the encounter
parameters as well as on the characteristics of the host
potential. Clearly, the ultimate goal would be to infer the properties of the
perturbers from the analysis of the gaps observed.

\cite{Erkal2015} developed a framework that predicts the evolution of gaps formed in streams that are orbiting on circular orbits. Using this model, \cite{Erkal2015b} showed how to infer the properties of a subhalo from the properties of a gap, down to a degeneracy in subhalo mass and relative velocity. A more recent model by \cite{Sanders2016DynamicsInteractions} focuses on modelling gaps in angle-frequency space, allowing for eccentric orbits \citep[see also][]{Bovy2017LinearSpectrum}. The authors validate several, but not all aspects of \cite{Erkal2015}, and argue for example that the velocity dispersions in the underlying stream affect the evolution of the gap, and thus should be taken into account. A caveat of all these models is that they are not fully analytical - and thus always rely on numerical exploration of the parameter space - or they are limited to circular orbits only. For this reason, we presented a fully analytical model for the evolution of gaps in streams \citep[HK16 hereafter]{Helmi2016} orbiting in spherical potentials.

In this work, we extend the HK16 model to streams orbiting in axisymmetric potentials. The model presented here not only predicts the behaviour of the size of the gap as a function of time but also its central density and their dependence on the characteristic parameters of the encounter. This paper is structured as follows. In Sec.~\ref{sec:methods}, we describe the model in detail and its predictions for the properties of gaps. In Sec.~\ref{sec:results} we validate our model with N-body experiments. Subsequently, in Sec.~\ref{sec:observables} we analyse the dependencies of the gap's properties on the collision parameters and investigate possible degeneracies in the parameters. Finally, we present a discussion and conclusions in Sec.~\ref{sec:CandD}.

\section{Methods}\label{sec:methods}

The main reason to extend our HK16 model, which only works for spherical potentials, is that the Milky Way is more realistically described as an axisymmetric system. From a dynamical point of view, breaking the spherical symmetry will add a degree of freedom to the system. 

The notation we use here is very similar to that employed in HK16. It builds on the action-angle stream description of \citet[][HW99 hereafter]{Helmi1999a}, see also \citet{Helmi2007}.

\subsection{Choice of the potential}\label{sec:potential}

We are somewhat restricted in our choice for a potential for the Milky Way because our approach is based on the use of action-angle variables. These can only be calculated in potentials that are separable in the coordinates. For this reason, we will use St\"ackel potentials, which are separable in ellipsoidal coordinates and are fully integrable (in fact, they are the only type of potentials with this property).

Because the (inner part of the) Milky Way is best described as an oblate system, we will use a set of prolate spheroidal coordinates $(\lambda,\phi,\nu)$ which we adopt from \cite{deZeeuw1985EllipticalPotentials}. The coordinate $\phi$ is the azimuthal angle and the other two coordinates, $\lambda$ and $\nu$, are the roots for $\tau$ in
\begin{equation}
 \frac{R^2}{\tau + \alpha} + \frac{z^2}{\tau + \gamma} = 1,
 \label{eq:spheroidal_coords}
\end{equation}
where $R = x^2 + y^2$, and $\alpha$ and $\gamma$ are constants related to the shape of the spheroid. The most general form of a St\"ackel potential in these coordinates is
\begin{equation}
 \Phi(\lambda,\nu) = \frac{(\nu + \gamma)G(\nu) - (\lambda+\gamma)G(\lambda)}{\lambda - \nu},
\end{equation}
where $G(\tau)$ determines the exact shape of the potential. For $G(\tau)$ we choose a two-component Kuzmin-Kutuzov potential, which takes the following form
\begin{equation}
 G(\tau) = \frac{GM_h}{\sqrt{\tau} + c_h} + \frac{GM_d}{\sqrt{\tau-q} + c_d},
\end{equation}
where $q$ is a parameter set by the choice of the different axis ratios for the components taking into account the constraint that the sum remains a St\"ackel potential: $\lambda_h - \nu_h = \lambda_d - \nu_d$, or $\lambda_d = \lambda_h - q$ and $\nu_d = \nu_h - q$, where
\begin{equation}
q = c_h^2\frac{\epsilon_h^2 - \epsilon_d^2}{1-\epsilon_d^2}, \qquad {\rm with}~q \ge 0.
\end{equation}
Here the ratio of the semi-major $a$ and semi-minor $c$ axis $\epsilon^2 = \alpha / \gamma$ (i.e. the flattening of the system) is a free parameter for each component, where $\alpha = -a^2$ and $\gamma = -c^2$. Finally, we define the fraction of the mass of the disc with respect to the total mass as $k = M_d/M_{\rm tot}$, with $M_{\rm tot} = M_d + M_h$. We recommend the interested reader to consult \cite{Dejonghe1988} for more details on axisymmetric St\"ackel potentials.

\begin{figure}
  \centering
    \includegraphics[width=0.9\hsize]{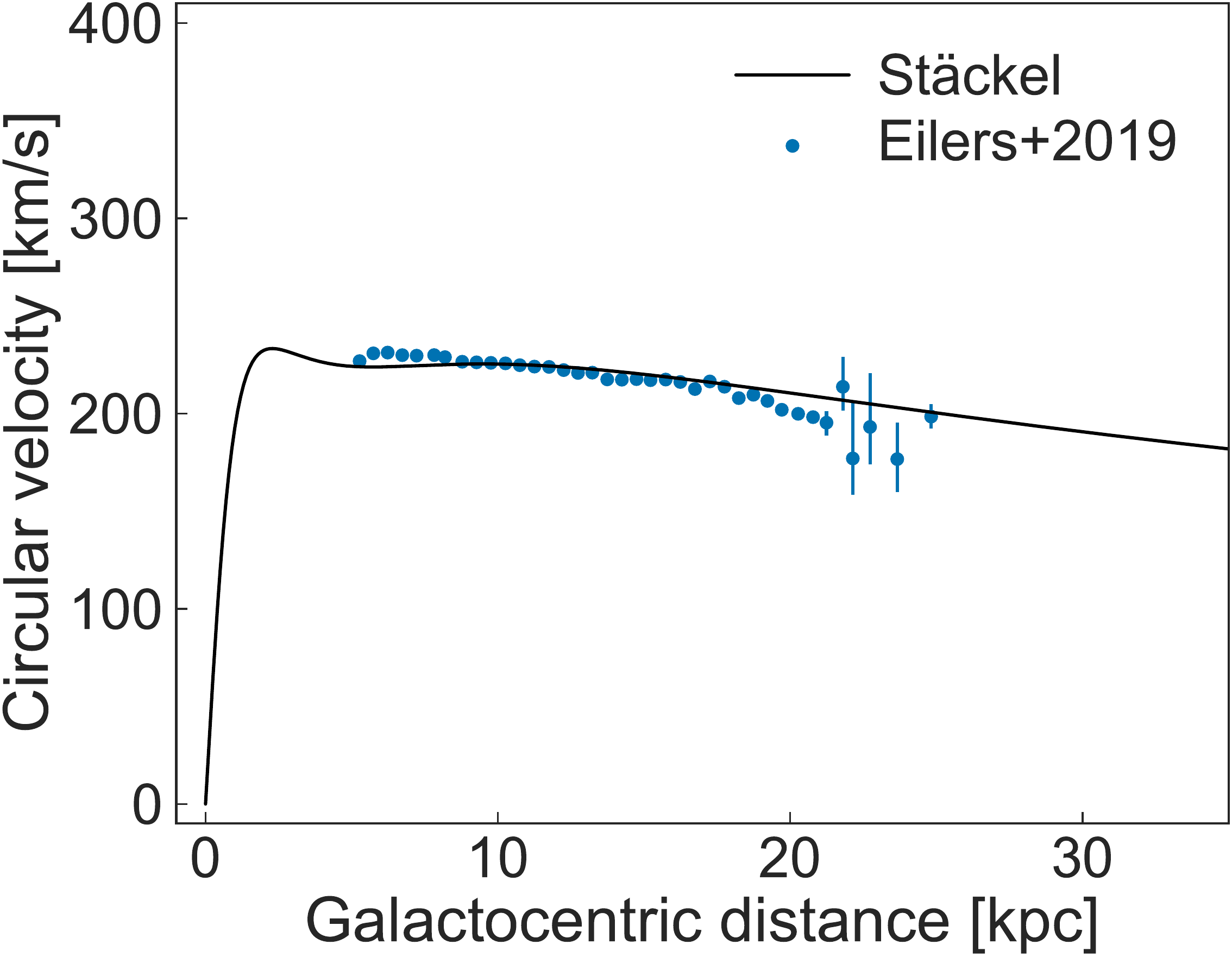}
  \caption{Circular velocity in the plane of the disc of our Milky Way model. The St\"ackel potential comprises a disc and a halo component and realistically describes the circular velocity of the Milky Way in the inner $\sim 20$ kpc as can be seen by comparing
to the determinations by \cite{Eilers2019TheKpc}.}
  \label{fig:vcurves}
\end{figure}

The resulting potential is therefore described by five parameters, namely the total mass $M_\mathrm{tot}$, the fraction of mass in the disc $k$, the scale length of the halo component $a_h$, and the flattening parameters of the halo $\epsilon_h$ and disc $\epsilon_d$. Here we set these parameter values to:
$M_\mathrm{tot} = 4.0 \cdot 10^{11}~\mathrm{M_\odot}$,
$k = 0.11$,
$a_h = 7.0$~kpc,
$\epsilon_h = 1.02$,
$\epsilon_d = 75.0$ \citep[which are based on][interested readers might want to consult also \citealt{Reino2020GalacticData}, where two-component St\"ackel potentials are fit to several streams around the Milky Way using {\it Gaia} DR2]{Batsleer1994,Famaey2003Three-componentParameters}. The resulting potential matches reasonably well the circular velocity curve of the Milky Way, as can be seen from Fig.~\ref{fig:vcurves} (solid black line). This can be inferred by comparison to the recently estimated circular velocity curve from \cite{Eilers2019TheKpc} (in blue).  

\subsection{Impulse approximation}\label{sec:impulse}

\begin{figure}
  \centering
    \includegraphics[width=\hsize]{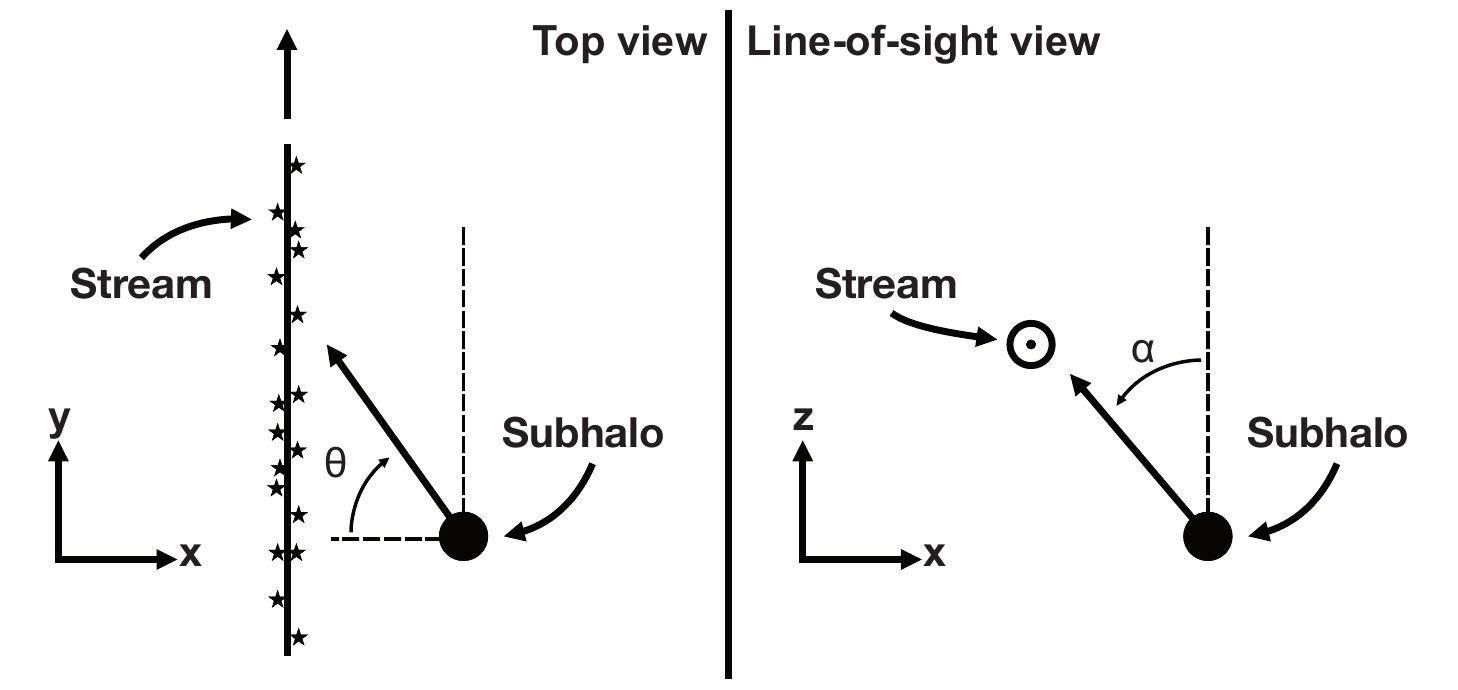}
  \caption{Schematic overview of the stream-subhalo interaction. Left: view seen from the top. Right: view seen along the line of sight (stream moves into the paper).}
  \label{fig:illustration}
\end{figure}

Before diving into the model, we will first describe the impact that a subhalo has on a cold stream. The gravitational interaction of a subhalo is well described by the impulse approximation\footnote{see Sec.~8.2 from \cite{Binney2008GalacticDynamics}.} \citep{Yoon2011ClumpyStructures, Carlberg2013}.
We define a reference system where the stream is aligned along the $y$-axis, and moves in the positive $y$-direction \citep[similar to the system of][c.f. their Fig.~2]{Erkal2015}. In this co-moving frame the relative velocity vector of the subhalo is $\boldsymbol w =  w(-\cos{\theta}\sin{\alpha},\sin{\theta},\cos{\theta}\cos{\alpha})$, or $\boldsymbol w = (-w_\perp\sin\alpha, w_\parallel,w_\perp\cos\alpha)$, where $w_\perp = w\cos\theta$ and $w_\parallel = w\sin\theta$. Figure~\ref{fig:illustration} illustrates the geometry of the stream-subhalo encounter.

The change of velocity (i.e. the impulse) of a particle along the stream due to the encounter is
\begin{equation}
 \Delta v_i = \int^\infty_{-\infty } a_i(\boldsymbol{x}, \boldsymbol{w}, M, r_s) \mathrm{d}t,
 \label{eq:impulsapproximationintegral}
\end{equation}
where $i = (x,y,z)$. The acceleration $a_i$ is a function of the relative velocity $\boldsymbol w$, the distance to the point of impact $y$, and of the subhalo mass $M$ and scale radius $r_s$. 
We model the subhalos as Plummer spheres but the expressions can be generalised for other profiles \citep{Sanders2016DynamicsInteractions}. The change in velocities in all three coordinates at the time of the impulse according to Eq.~\eqref{eq:impulsapproximationintegral} is
\begin{subequations}
\begin{equation}
 \frac{\Delta v_x}{2GM} = 
 \frac{yw_{\perp}w_{\parallel}\sin{\alpha}}{w\bigg(r_s^2w^2 + y^2w_{\perp}^2\bigg)} =
 \frac{y\cos{\theta}\sin{\theta}\sin{\alpha}}{w\bigg(r_s^2 + y^2\cos^2{\theta}\bigg)},
 \label{eq:EB_dvx}
\end{equation}
\begin{equation}
 \frac{\Delta v_y}{2GM} = 
 -\frac{w_{\perp}^2y}{w\bigg(r_s^2w^2 + y^2w_{\perp}^2\bigg)} = 
 -\frac{y\cos^2{\theta}}{w\bigg(r_s^2 + y^2\cos^2{\theta}\bigg)},
 \label{eq:EB_dvy}
\end{equation}
\begin{equation}
 \frac{\Delta v_z}{2GM} = 
 -\frac{yw_{\perp}w_{\parallel}\cos{\alpha}}{w\bigg(r_s^2w^2 + y^2w_{\perp}^2\bigg)} = 
 -\frac{y\cos{\theta}\sin{\theta}\cos{\alpha}}{w\bigg(r_s^2 + y^2\cos^2{\theta}\bigg)}.
 \label{eq:EB_dvz} 
\end{equation}
 \label{eq:Dv_equations}
\end{subequations}
The above expressions are valid for direct encounters, that is when the impact parameter $b=0$. Eqs.~(1-3) in \cite{Erkal2015} provide a more general form for the velocity changes which take into account the parameter $b$. This parameter enters into the equations above through $r_s^2 \rightarrow r_s^2+b^2$. 

We assume that the stream is linear over the scale where the impulse is significant. Moreover, the equations above assume that that the stream is a 1D-structure. This approximation is sufficient when the width of the stream is smaller than the scale radius of the subhalo. However, the expressions can be generalised to the full 3D case, for which we find
\begin{equation}
 \Delta v_i(\boldsymbol{x}) = -\frac{2GM}{w}\frac{w^2x_i - w_i(\boldsymbol{x}\cdot\boldsymbol{w})}{(r_s^2 + \boldsymbol{x}\cdot{\boldsymbol{x}})w^2 - (\boldsymbol{x}\cdot\boldsymbol{w})^2},
 \label{eq:3Dv_equations}
\end{equation}
with $i=(x,y,z)$. To gain insight into the model we will use the equations of the 1D approximation in this section. However, when evaluating the model we will use the full 3D equations.

From Eq.~\eqref{eq:Dv_equations} we can find the maximum kick in velocities $\Delta v_i^{\rm max}$ and at what distance $y_{\rm max}$ to the centre of impact it occurs,
\begin{subequations}
\begin{equation}
 \Delta v^\mathrm{max}_i = -\frac{2GM}{w}\frac{w^2x_i - w_iw_yy_\mathrm{max}}{(r_s^2 + y_\mathrm{max}^2)w^2 - (y_\mathrm{max}w_y)^2},
 \label{eq:maxkickvel}
\end{equation}
where $x_i = [0,y_\mathrm{max},0]$ and
\begin{equation}
 y_\mathrm{max} =\frac{wr_s}{\sqrt{w^2 - w_y^2}} = \frac{r_s}{\cos{\theta}}.
 \label{eq:maxkickpos}
\end{equation}
\label{eq:maxkicks}
\end{subequations}
Typical profiles of $\Delta v_y(y)$ are shown and discussed in Sec.~\ref{sec:results}, see Fig.~\ref{fig:kickprofile_configurations}.

\subsection{Action-Angle variables}\label{sec:aavar}
This section aims to serve as a brief introduction to these variables, and it is by no means exhaustive or comprehensive. For more details on action-angle variables, the reader could consult \cite{Goldstein2002ClassicalMechanics, Binney2008GalacticDynamics}.

Orbits in smooth and simple potentials (e.g. spherical, axisymmetric, triaxial) have a number of integrals of motion: properties that do not change in time and serve to characterise them. For a spherical potential, the integrals of motion are the total energy (or the Hamiltonian) and the angular momentum vector. Orbits in axisymmetric systems (e.g. disc galaxies) typically have up to three integrals of motion: the total energy, the momentum in the azimuthal direction, and a non-classical integral which in most cases does not take an analytic form. 

For separable potentials (e.g. the St\"ackel potentials discussed in Sec.~\ref{sec:potential}) there exist three isolating integrals ${\bf J}$, known as the actions. Each action is paired with a conjugate coordinate $\boldsymbol{\Theta}$, the angles. Together, these coordinates make up the action-angle variables $(\boldsymbol{\Theta},{\bf J})$. The actions uniquely define the orbit, that is, a point in action-space corresponds to a complete orbit in phase-space. The conjugate angles define the phase, that is they specify where along the orbit a body is located at any given time.

To obtain the action-angle variables we make use of the Hamiltonian $H$, which being an integral of motion must depend on the actions (i.e. $H = H({\bf J})$). The rate of change of the angles ${\boldsymbol{\dot \Theta}} = \partial H/ \partial {\bf J}$ is known as the frequency  $\boldsymbol{\Omega}({\bf J})$. Therefore 
\begin{equation}
    \boldsymbol{\Theta}(t) = \boldsymbol{\Theta}_0 + \boldsymbol{\Omega} \,t,
    \label{eq:angeltimedep}
\end{equation}
and hence the angles are linearly dependent on time. Finally, the actions ${\bf J}$ of a bound orbit in a separable potential are defined as
\begin{equation}
    {\bf J} = \frac{1}{2\pi}\oint {\bf p} \cdot {\rm d} {\bf q},
\end{equation}
where $({\bf q,p})$ are any set of generalised phase-space coordinates and momenta.

\subsection{Size of the gap using an actions-angles framework} \label{sec:AA2orb}

The analytical framework of the method that we will use to describe the evolution of a gap in a stream with time, was first established by HW99. Originally, this framework was used to describe
the divergence in the orbits of a distribution of nearby particles. It
makes use of a linearised Taylor expansion around a central
orbit. In our case, we will model the size of the gap as the spatial
separation of two orbits: one on each side of the gap. These orbits
are taken to be those of the particles that receive the largest
impulse from the subhalo flyby. In practice, this is equivalent to
modelling the (size of the) gap as twice the separation of the central
orbit and one of the edges of the gap, as gaps are symmetric with respect to their centre.

\subsubsection{Generalities}

Let us consider a central
orbit and some other orbit separated by $\Delta {\bf X}_0$ and
$\Delta {\bf V}_0$, where the subscript is used to denote the time of the impact between the subhalo and the stream, 
$t=t_0$. To calculate the evolution of this separation vector we first
transform it to action-angle variables
\begin{equation}
 \begin{bmatrix}
  \Delta \boldsymbol\Theta_0 \\[0.3em]
  \Delta \boldsymbol{\mathrm{J}}_0
 \end{bmatrix} = {\cal M}_0
 \begin{bmatrix}
  \Delta \boldsymbol{\mathrm{X}}_0 \\[0.3em]
  \Delta \boldsymbol{\mathrm{V}}_0
 \end{bmatrix},
 \label{eq:xyz2AA}
\end{equation}
where ${\cal M}_0$ is a matrix calculated at $t=t_0$ that locally transforms from Cartesian coordinates to action-angles. In practice, the transformation is a product of matrices
\begin{equation}
  {\cal M}_0 = {\cal M}^{\mathrm{AA} \leftarrow \mathrm{st}}_0 {\cal M}^{\mathrm{st} \leftarrow \mathrm{cyl}}_0 {\cal M}^{\mathrm{cyl} \leftarrow \mathrm{xyz}}_0,
\label{eq:tmatrix}
\end{equation}
where ${\cal M}_0^{\beta \leftarrow \alpha}$ transforms the set of coordinates $\alpha$ to the set $\beta$, $_{\mathrm{xyz}}$ indicating Cartesian coordinates, $_{\mathrm{cyl}}$ cylindrical coordinates, $_{\mathrm{st}}$ spheroidal coordinates used for the St\"ackel potential, and $_{AA}$ action-angle variables. 

Next, the separation vector in action-angles can be evolved in time by expanding linearly Eq.~\eqref{eq:angeltimedep} and making use of the matrix $\Omega'$ 
\begin{equation}
 \Omega' =  \begin{bmatrix}
            I_3 & \partial\boldsymbol\Omega/\partial\boldsymbol{\mathrm{J}}\,t \\[0.3em]
            0 & I_3
           \end{bmatrix},
\label{eq:omega-prime}
\end{equation}
At any point in time, the separation in action-angle coordinates can be transformed back to Cartesian coordinates locally, and therefore 
\begin{equation}
  \begin{bmatrix}
  \Delta \boldsymbol{X}_t \\[0.3em]
  \Delta \boldsymbol{V}_t
 \end{bmatrix} = {\cal M}^{-1}_t\Omega' {\cal M}_0
 \begin{bmatrix}
  \Delta \boldsymbol{\mathrm{X}}_0 \\[0.3em]
  \Delta \boldsymbol{\mathrm{V}}_0
 \end{bmatrix},
 \label{eq:xyz02xyzt}
\end{equation}
where ${\cal M}^{-1}_t$ is the (local) transformation back to Cartesian coordinates at time $t$ and at the location of the central orbit of the gap. 

Finally, the size of the gap can be taken as twice the separation calculated in Eq.~\eqref{eq:xyz02xyzt}. The initial separation of the two orbits describing the gap can be obtained assuming Eq.~\eqref{eq:maxkickvel} and Eq.~\eqref{eq:maxkickpos}. Since the two orbits are typically separated a few kpc initially, we need to add the velocity gradient of the orbit to the separation in velocities, so $\Delta V_0 = \Delta v^\mathrm{max} + \delta v_\mathrm{orbit}$. We note that this is an {\it ad hoc} fix to the non-local nature (finite extent) of the stream. It takes into account that the velocity of the stream particles changes as a function of location.

\subsubsection{Long-term behaviour}

The growth rate of the size of the gap can be derived from Eq.~\eqref{eq:xyz02xyzt} in a similar fashion as shown in HK16. In the limit where $t\gg t_0$ (or better $t/t_{\rm orb} \gg 1$), this equation simplifies to
\begin{equation}
  \begin{bmatrix}
  \Delta \boldsymbol{X}_t \\[0.3em]
  \Delta \boldsymbol{V}_t
 \end{bmatrix} \sim t \begin{bmatrix}
 {\cal M}^{-1}_{t,1}\partial\boldsymbol\Omega/\partial\boldsymbol{\mathrm{J}}\Delta\boldsymbol{\mathrm{J}}_0 \\[0.3em]
 {\cal M}^{-1}_{t,3}\partial\boldsymbol\Omega/\partial\boldsymbol{\mathrm{J}}\Delta\boldsymbol{\mathrm{J}}_0
 \end{bmatrix},
 \label{eq:workingout_prediction_2}
\end{equation}
where ${\cal M}^{-1}_{t,1}$ is the upper left submatrix of ${\cal M}^{-1}_{t}$, and ${\cal M}^{-1}_{t,3}$ is the bottom left submatrix.
The spatial separation of the two orbits is equal to the length of vector $\Delta \boldsymbol{X}_t$
\begin{equation}
 |\Delta \boldsymbol{X}_t| = \sqrt{\Delta \boldsymbol{X}_t^\dagger\Delta \boldsymbol{X}_t} \sim t\sqrt{\Delta\boldsymbol{\mathrm{J}}_0^\dagger f_{\boldsymbol x, \boldsymbol \Omega}\Delta\boldsymbol{\mathrm{J}}_0}.
 \label{eq:Dx_pred}
\end{equation}
In this equation 
$f_{\boldsymbol x, \boldsymbol \Omega} = \partial\boldsymbol\Omega/\partial\boldsymbol{\mathrm{J}}\,
{{\cal M}^{-1}_{t,1}}^\dagger {\cal M}^{-1}_{t,1}\,
\partial\boldsymbol\Omega/\partial\boldsymbol{\mathrm{J}}$.
Similarly, we can calculate the velocity difference of the two orbits $\Delta \boldsymbol V$
\begin{equation}
 |\Delta \boldsymbol{V}_t| = \sqrt{\Delta \boldsymbol{V}_t^\dagger\Delta \boldsymbol{V}_t} \sim t\sqrt{\Delta\boldsymbol{\mathrm{J}}_0^\dagger f_{\boldsymbol v, \boldsymbol \Omega}\Delta\boldsymbol{\mathrm{J}}_0},
 \label{eq:Dv_pred}
\end{equation}
where $f_{\boldsymbol v, \boldsymbol \Omega} =
\partial\boldsymbol\Omega/\partial\boldsymbol{\mathrm{J}}\,
{{\cal M}^{-1}_{t,3}}^\dagger {\cal M}^{-1}_{t,3}\,
\partial\boldsymbol\Omega/\partial\boldsymbol{\mathrm{J}}$.

We note that the terms $f_{\boldsymbol x, \boldsymbol \Omega}$ and $f_{\boldsymbol v, \boldsymbol \Omega}$ are dependent on the orbit of the gap and its location, but they do not depend on the impact parameters. Both $|\Delta \boldsymbol{X}_t|$ and $|\Delta \boldsymbol{V}_t|$ are \emph{linearly} dependent on time $t$, similar to gaps orbiting in spherical potentials. Interestingly, the ratio of the two separations is constant with time - which potentially can be used to infer the properties of the gap at any time (as we will demonstrate also in Sec.~\ref{sec:observables}).

\subsection{Density of stream gaps}\label{sec:centraldensity}

We now build further on the framework developed by HW99 and focus on
modelling the evolution of the density of the gap. The impulse
imparted on the stream by the subhalo increases the local velocity
dispersion of the stars in the gap. This causes it to grow faster
and thus appear as under dense region in comparison to the
neighbouring parts of the stream. If we know the central orbit of the
gap and the initial phase-space distribution around it, we
can calculate the evolution of the density in the gap. 

\subsubsection{Generalities}

We will describe this initial
phase-space distribution as a multi-variate
Gaussian distribution, in other words
\begin{equation}
	f(\boldsymbol{x},\boldsymbol{v}) = f_0\exp{\bigg(-\frac{1}{2}\Delta^{\dagger}_{\varpi,0} \sigma_{\varpi,0}   \Delta_{\varpi,0}\bigg)},
    \label{eq:distrfunct0}
\end{equation}
where $f_0$ is the phase-space density at $t=t_0$, $\Delta_{\varpi,0}$ is a separation vector: $\Delta_{\varpi,0} = \xi_i - \xi_{c,i}$ and where $\xi_i = [x,y,z,v_x,v_y,v_z]$ and $\xi_{c,i}$ is the central point of the distribution (which we will take to be the location where the subhalo impacts the stream, or in the terminology previously used, the central orbit) at $t=t_0$. The matrix $\sigma_{\varpi,0}$ is the inverse of the covariance matrix of the phase-space coordinates.

To compute the initial dispersion matrix of the gap, we start from the original, unperturbed distribution and add the impulse in the velocities according to Eq.~\eqref{eq:3Dv_equations}. That is, we transform $\sigma^{\mathrm{stream}}_{\varpi,0} + {\rm impulse} \rightarrow \sigma^{\mathrm{gap}}_{\varpi,0}$. Below we show how to calculate the new covariance matrix in the regime where the stream is approximated by a 1D-structure, but in Appendix~\ref{sec:3D_cov} we provide the full 3D expressions. 

The most general form of the initial unperturbed covariance matrix $\Sigma_{\varpi,0}$ is
\begin{equation}
	\Sigma_{\varpi,0} = \sigma_{\varpi,0}^{-1} =  \begin{pmatrix}
  \sigma_x^2 & C(x,y)  & \cdots\\
  C(y,x) & \sigma_y^2 & \cdots \\
  \vdots  & \vdots  & \ddots \\
 \end{pmatrix},
	\label{eq:covariance}
\end{equation}
where $C(x,y)$ is the covariance of $x$ and $y$ and $\sigma_x$ is the standard deviation of $x$. The covariance matrix can be represented with $3\times3$ block matrices

\begin{equation}
\Sigma_{\varpi,0} =
\begin{pmatrix}
  \mathbb{C}_{\boldsymbol{x},\boldsymbol{x}} & \mathbb{C}_{\boldsymbol{x},\boldsymbol{v}}\\
  \mathbb{C}_{\boldsymbol{v},\boldsymbol{x}} & \mathbb{C}_{\boldsymbol{v},\boldsymbol{v}}
 \end{pmatrix}, \qquad {\rm where}\quad\mathbb{C}_{\boldsymbol{x},\boldsymbol{v}} = \mathbb{C}_{\boldsymbol{v},\boldsymbol{x}}^{\dagger}.
\label{eq:bigSigma_0}
\end{equation}

We now proceed to compute the perturbed covariance matrix by computing the changes of each individual element due to the encounter with the subhalo. The impulse only affects the velocities. Therefore, the position block matrix ($\mathbb{C}_{\boldsymbol{x},\boldsymbol{x}}$) does not change during the encounter. The first element with a velocity term is in the $\mathbb{C}_{\boldsymbol{x},\boldsymbol{v}}$ block matrix
\begin{equation}
C(v_x,x) = \frac{1}{n} \sum_{i=1}^{n}(v_{x_i} - \mu_{v_x})(x_i - \mu_{x}),
\end{equation}
where $n$ is the total number of particles, $\mu_{v_x}$ and $\mu_{x}$ are the mean $v_{x}$ and $x$ of the distribution in the region around the gap. After applying the impulse, the new covariance element becomes
\begin{equation}
C(v_x + \Delta v_x,x) = \frac{1}{n} \sum_{i=1}^{n}(v_{x_i} + \Delta v_{x_i} - \mu_{v_x} - \Delta\mu_{v_x})(x_i - \mu_{x}),
\end{equation}
where $\Delta v_{x_i}$ is the velocity change of particle $i$, and $\Delta\mu_{v_x}$ is the shift of the mean velocity of all particles.
Since the kicks are symmetric around the central point, the mean shift of velocities is zero $\Delta\mu_{v_x} = 0$, and we can rewrite the covariance term as
\begin{equation}
C(v_x + \Delta v_x,x) = C(v_x,x) + \frac{1}{n} \sum_{i=1}^{n}\Delta v_{x_i} (x_i - \mu_{x}).
\label{eq:covdecomposition}
\end{equation}
Considering that the covariance matrix describes the central density (i.e. positions close to the centre) we can express the kick $\Delta v_x$ as a function that is only linearly dependent on $y$, since the quadratic term in the denominator of Eq.~\eqref{eq:EB_dvx} is negligible (i.e. $r_s^2w^2 >> y^2w_{\perp}^2$). Moreover, since the velocity kicks are calculated in a frame where there is symmetry with respect to $y$ (i.e. $\mu_{y} = 0$) we can rewrite the last term in the equation above as
\begin{equation}
   \frac{1}{n} \sum_{i=1}^{n}\Delta v_{x_i} (x_i - \mu_{x}) =  2GM\frac{w_{\perp}w_{\parallel}\sin{\alpha}}{r_s^2w^3} \frac{1}{n} \sum_{i=1}^{n} (y_i - \mu_{y}) (x_i - \mu_{x}), 
\end{equation}
which is equal to
\begin{equation}
    C(v_x + \Delta v_x,x) = C(v_x,x) + 2GM\frac{w_{\perp}w_{\parallel}\sin{\alpha}}{r_s^2w^3} C(y,x).
\end{equation}
The new covariance term $C(v_x + \Delta v_x,x)$ can be expressed as the old covariance term plus a new term that depends on the impact parameters. The procedure shown above can be extended to all covariance terms of the form $C(\alpha,v_\beta + \Delta v_\beta)$ and $C(v_\beta + \Delta v_\beta, \alpha)$, where $(\alpha,\beta) = (x,y,z)$.

Using similar arguments, it is easy to show that covariance terms in the velocity submatrix ($\mathbb{C}_{\boldsymbol{v},\boldsymbol{v}}$) take the following general form
\begin{align}
    C(v_\alpha + \Delta v_\alpha, v_\beta + \Delta v_\beta) = 
    &C(v_\alpha,v_\beta) + C(\Delta v_\alpha, v_\beta) \\
    &+ C(v_\alpha, \Delta v_\beta) +
    C(\Delta v_\alpha, \Delta v_\beta).\nonumber
\end{align}
For example, for $\alpha=x$ and $\beta=y$, and following similar procedures as above
\begin{align}
&C(\Delta v_x, v_y) = 2GM\frac{w_{\perp}w_{\parallel}\sin{\alpha}}{r_s^2w^3}C(y,v_y), \\
&C(v_x, \Delta v_y) = -2GM\frac{w_{\perp}^2}{r_s^2w^3}C(v_x,y), \\
&C(\Delta v_x, \Delta v_y) = -\bigg(2\frac{GM}{r_s^2w^3}\bigg)^2w_{\perp}^3w_{\parallel}\sin{\alpha}\,C(y,y).
\end{align}

\begin{figure}
  \centering
    \includegraphics[trim=0cm 0cm 3cm 2.5cm, clip,width=\hsize]{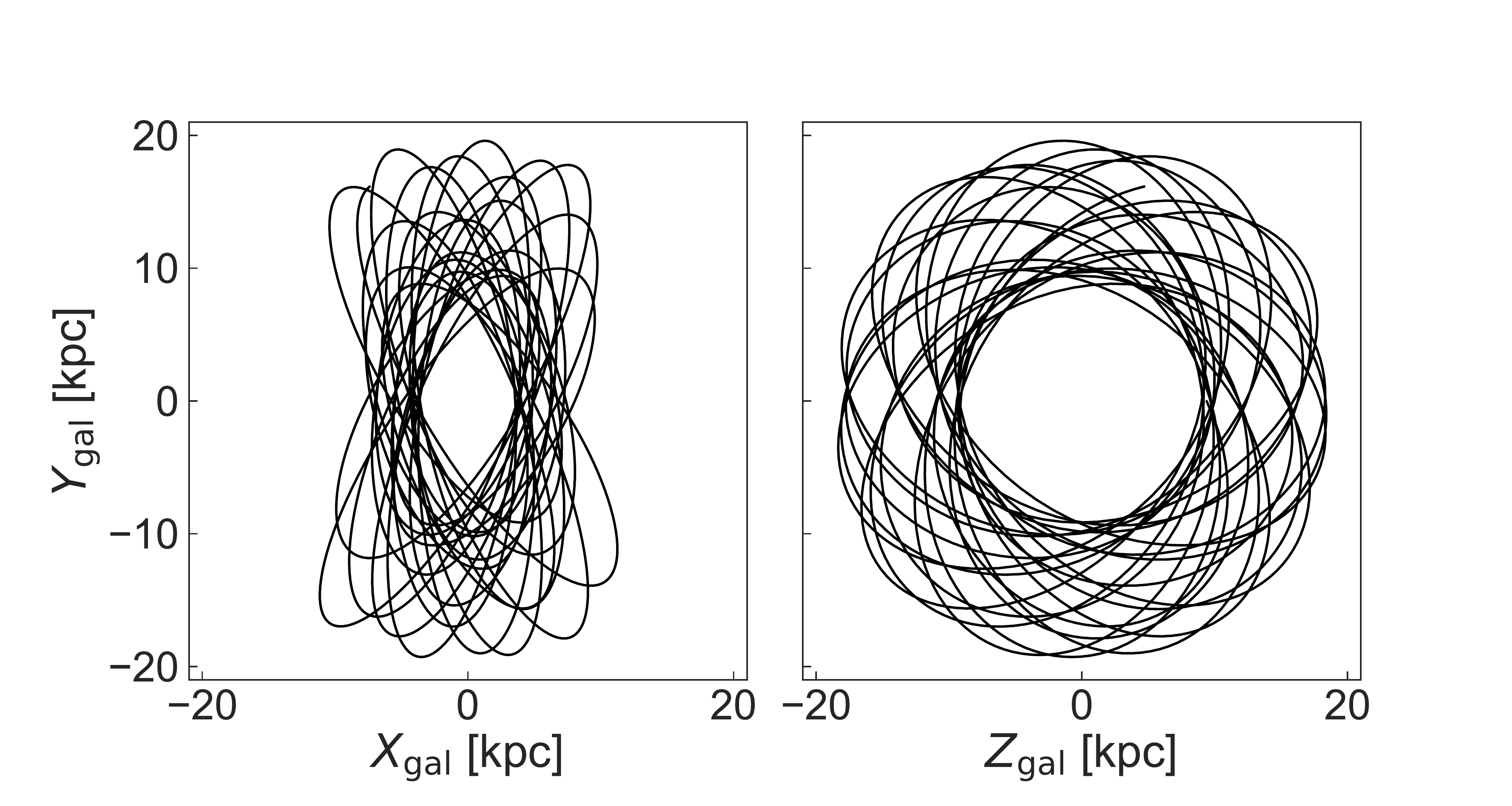}
    \includegraphics[trim=0cm 0cm 3cm 2.5cm, clip,width=\hsize]{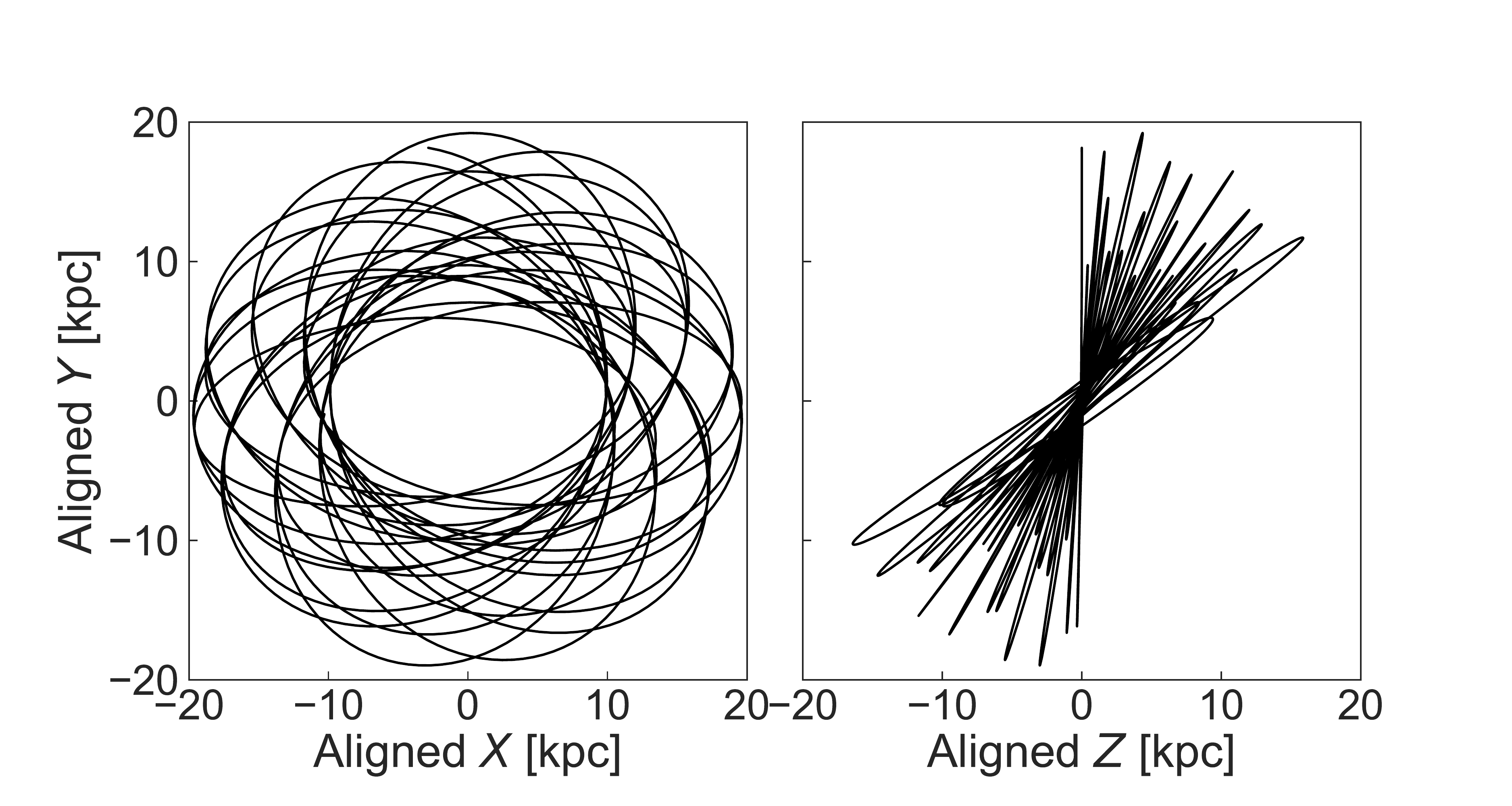}
  \caption{The orbit of the stream shown in galactocentric Cartesian coordinates, evolved for 10 Gyr forward in time in the St\"ackel axisymmetric Milky Way-like potential. Top: orbit in coordinates aligned with the host galaxy (e.g. plane of the disc is $z=0$). Bottom: coordinates aligned with the initial angular momentum vector of the orbit. The right bottom panel highlights the precession of the orbital plane, which is indicative of the non-spherical nature of the potential considered.}
  \label{fig:orbit}
\end{figure}

We now have full expressions for the matrix $\sigma_{\varpi,0}$ in Eq.~\eqref{eq:distrfunct0} representing the phase-space configuration around the gap at the time of the encounter $t=t_0$. By transforming $\sigma_{\varpi,0}$ to action-angle coordinates as
$\sigma_{\omega,0} = {{\cal M}^{-1}_0}^\dagger\sigma_{\varpi_0}{\cal M}^{-1}_0$, where ${\cal M}$ is
the transformation matrix defined in Eqs.~\eqref{eq:xyz2AA} and \eqref{eq:tmatrix}, we can calculate the evolution
in time of the covariance matrix in phase-space using Eq.~\eqref{eq:omega-prime}. 
This allows us to describe the local density of the portion of the stream around the location of the impact by the subhalo (i.e. of the gap) as  
\begin{equation}
	\rho(\boldsymbol{x}_c,t) = \int f(\boldsymbol{x}_c,\boldsymbol{v},t)\,\mathrm{d}^3\boldsymbol{v},
    \label{eq:density}
\end{equation}
where $\rho(\boldsymbol{x}_c,t)$ is the density of orbits in a location around the central orbit. In the principal axes, where the velocity covariance matrix is diagonal, this density takes a simple form 
\begin{equation}
	\rho(\boldsymbol{x}_c,t) = \rho_0/\sqrt{\det{|{\sigma}_v|}} \propto \rho_0 \sigma_{v_1}(t)\sigma_{v_2}(t)\sigma_{v_3}(t) ,
    \label{eq:centraldensity}
\end{equation}
where $\rho_0$ is the central density at $t=t_0$ and $\sigma_{v_1},\sigma_{v_2},\sigma_{v_3}$ are the velocity dispersions along the three principal axes.

\subsubsection{Long-term behaviour of the density}\label{sec:rhoevol}

Using the above formalism, it is possible to show that the density of a stream (and thus also that of a gap) decreases as a power law of time which depends on the number of degrees of freedom of the orbit of the stream \citep{Vogelsberger2008TheHaloes}
\begin{equation}
 \rho \propto t^{-n},\, \mathrm{ with \,\, n = d.o.f.}
\end{equation}
Ultimately these degrees of freedom are determined by the number of independent frequencies, and this number generally is dependent on the functional form of the potential. For axisymmetric galaxies the number of d.o.f. is 3 for most (non-resonant) orbits. On the other hand, for example, circular orbits only have one degree of freedom, implying that the density decreases much slower (i.e. $1/t$). 

HW99 derived a general expression for the central density at late times for streams (and gaps) in a general St\"ackel potential (see their Appendix~C) and found
\begin{equation}
 \rho({\bf x}_c,t) = \frac{\rho_0\,f_\mathrm{orb}}{\sqrt{\det{|\sigma_{\Theta_0}|}}} t^{-3},
 \label{eq:density_func_of_time}
\end{equation}
where $\rho_0$ is the initial density of the distribution,
$f_\mathrm{orb}$ is a constant determined by the central orbit, and
$\sigma_{\Theta_0}$ is the angle submatrix at $t=t_0$. This implies
that the ratio of the density of a perturbed to unperturbed stream is
a constant
\begin{equation}
 \delta\rho^\mathrm{gap}_\mathrm{str} = \sqrt{\frac{\det{|\sigma_{\Theta_0}|}_\mathrm{str}}{\det{|\sigma_{\Theta_0}|}_\mathrm{gap}}},
 \label{eq:dens_plateau}
\end{equation}
as all other variables are independent on the impact parameters. We will refer later in this work to this ratio of densities as the density contrast.

\subsection{Setting up the stream-subhalo encounter}\label{sec:Nbody}

To verify our model predictions, we perform N-body simulations of the encounter of a subhalo with a stream orbiting in the Milky Way potential described in Sec.~\ref{sec:potential}. To this end, we use a modified version of \mbox{\textsc{Gadget-2}} \citep{Springel2005TheGADGET-2}, where we model the host as the rigid potential, and the subhalo as a rigid Plummer sphere which is centred on a particle with a negligible mass that is put on a trajectory in the~host~potential.

The progenitor of the stream is modelled with $10^6$ test
particles\footnote{Because we use test particles, it is not strictly necessary to model their evolution using an N-body code such as {\tt Gadget}.} following a Gaussian distribution in 6D phase
space, with $\sigma_x = 0.2$ kpc and $\sigma_v = 0.5$ km/s. These
very low dispersions are chosen such that the stream has a high
density even a few Gyr after forming. In comparison, globular clusters
orbiting the Milky Way typically have a $\sigma_v$ of a few km/s
\cite[e.g.][2010 edition]{Harris1996}.

\begin{figure}
	\centering
    \includegraphics[trim={1cm 1cm 1cm 1cm}, clip, width=\hsize]{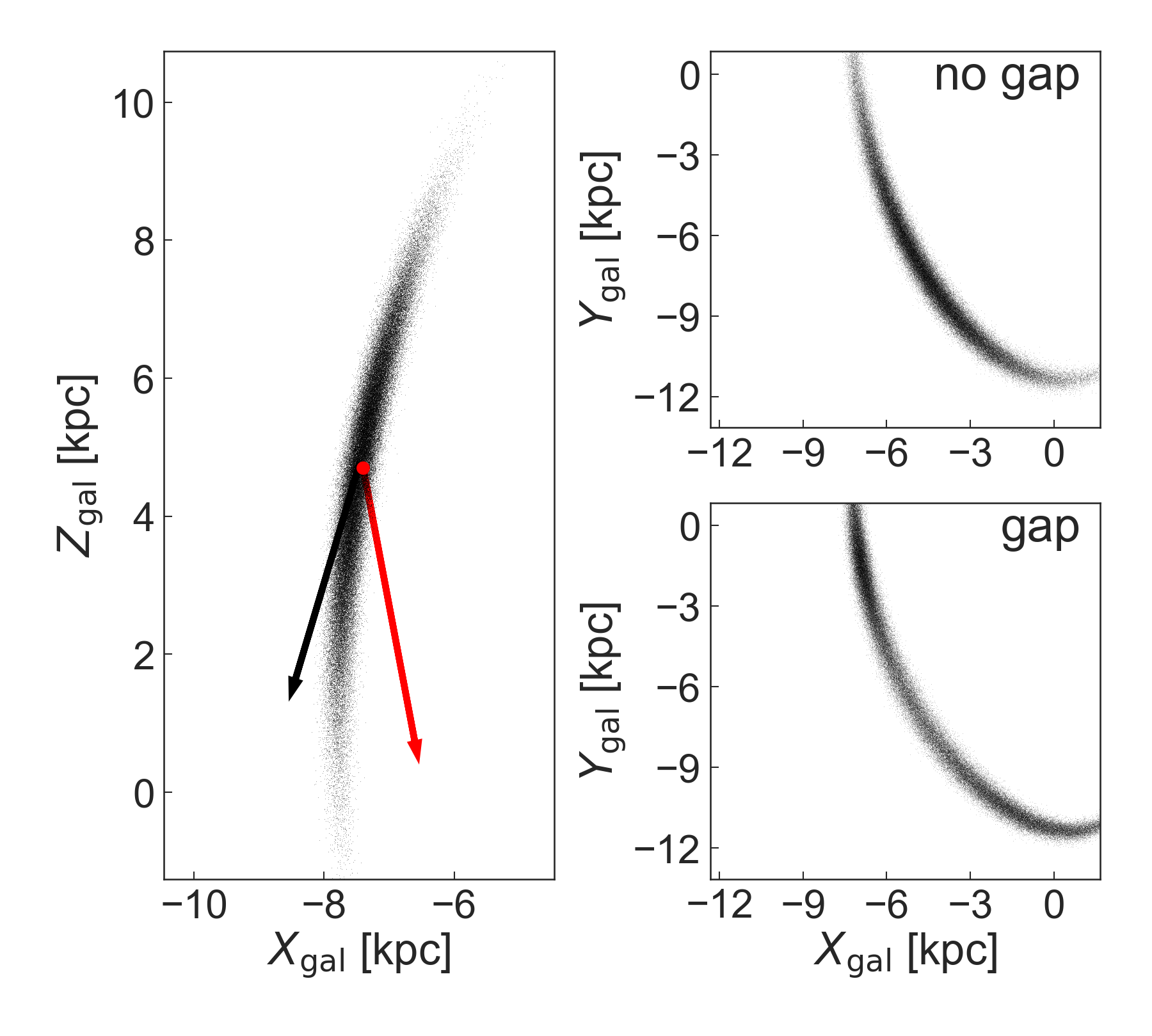}
    \caption{Left: Snapshot of a stream at the time of interaction with a subhalo. The stream is plotted with black dots and the centre of mass of the subhalo is shown as a red solid circle. The velocity vector of the stream is marked with a black arrow and that of the subhalo with a red arrow. Right: Both panels show the stream after 2~Gyr of evolution, 
in isolation in the top panel, and after the encounter with the subhalo in the bottom panel.}
    \label{fig:stream_subhalo_plots}
\end{figure}

\begin{table}
\caption{The masses and scale radii of the subhalos. The subhalos are modelled as rigid Plummer spheres.}
\label{table:subhalos}
\centering
\begin{tabular}{c c c c}
\hline\hline
 & subhalo 1 & subhalo 2 & subhalo 3   \rule{0pt}{2.3ex}\\
\hline
$\text{M}$ [$\text{M}_{\odot}$] & $10^{6}$ & $10^7$ & $10^{8}$ \rule{0pt}{2.3ex}\\
$\text{r}_{s}$ [kpc] & 0.35 & 0.59 & 1.35\\
\hline
\end{tabular}
\end{table}

The progenitor
is put on an elongated orbit with maximum distance from the centre
$r_\mathrm{max} \approx 20$ kpc and minimum distance
$r_\mathrm{min} \approx 10$ kpc, reaching $r_z = \pm 20$ kpc above the
plane of the disc, as shown in Fig.~\ref{fig:orbit}. After the
progenitor of the stream is evolved for $1$ Gyr in the host potential,
a subhalo is inserted on a trajectory that directly crosses the
stream. We remove the subhalo after the collision to isolate a single
interaction, and when its gravitational effect is sufficiently small
that it no longer affects the stream.

Fig.~\ref{fig:stream_subhalo_plots} shows an example of a stream-subhalo interaction. The left panel shows both the stream and a subhalo at the time of the collision. The right panels show a stream with and without an encounter, 2 Gyr after the interaction with the subhalo. The perturbed stream clearly shows a gap of several kpc in size at the centre of~the~panel.

\section{Results}\label{sec:results}

\begin{figure}
  \centering
  \includegraphics[width=\hsize]{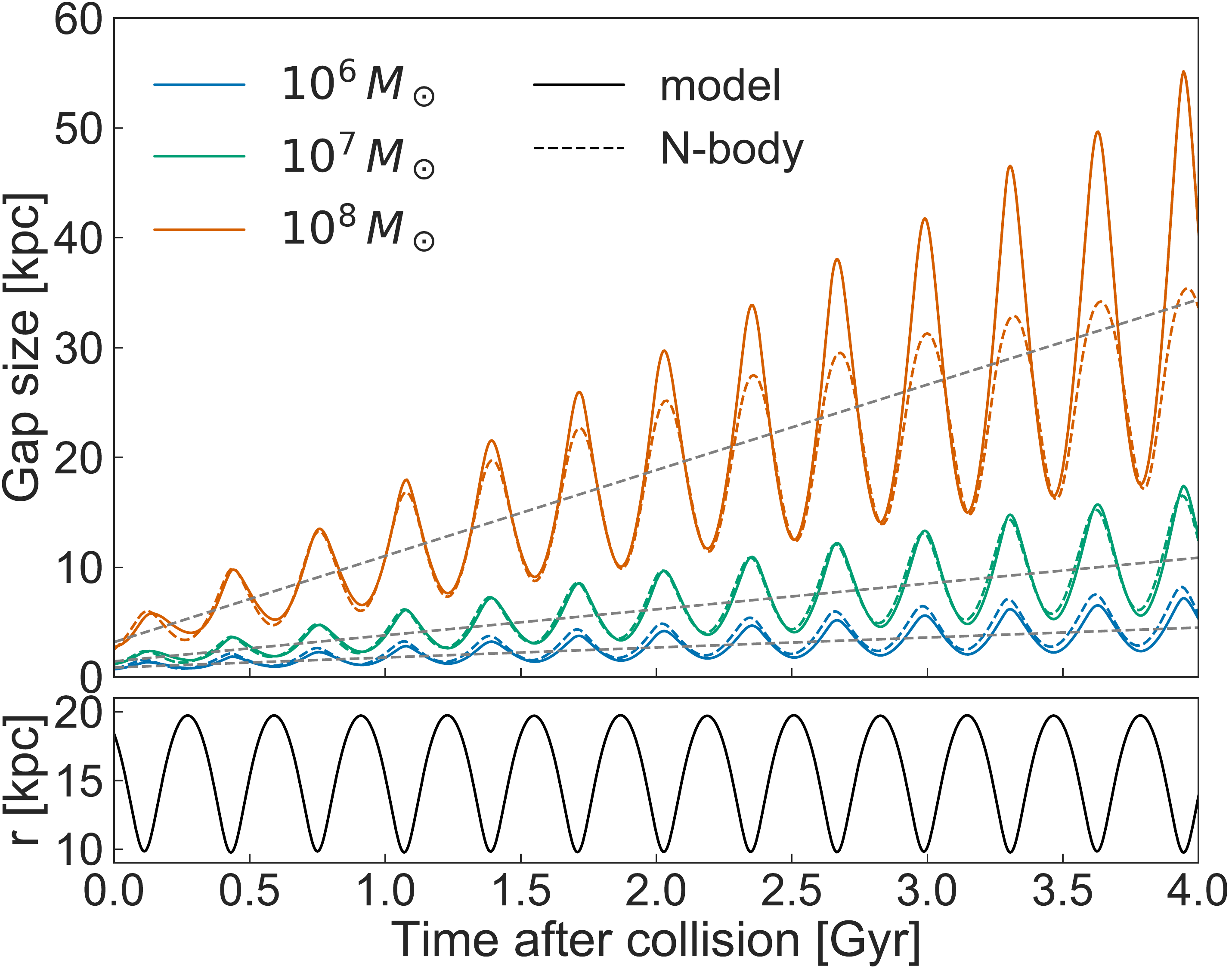}
  \caption{Top panel: Gaps growing in a stream orbiting an axisymmetric St\"ackel Milky Way-like potential. The lines with different colours show the size of the gaps induced by subhalos of different sizes. The solid curves correspond to our analytic model, while the dashed curves to gaps measured in the N-body experiments. The straight grey dashed lines illustrate the linear growth rate of the gaps. Bottom panel: Distance of the central orbit to the centre of the host potential.}
  \label{fig:modelvsnbody}
\end{figure}

\begin{figure}
  \centering
  \includegraphics[trim={1cm 1cm 1cm 1cm}, clip, width=\hsize]{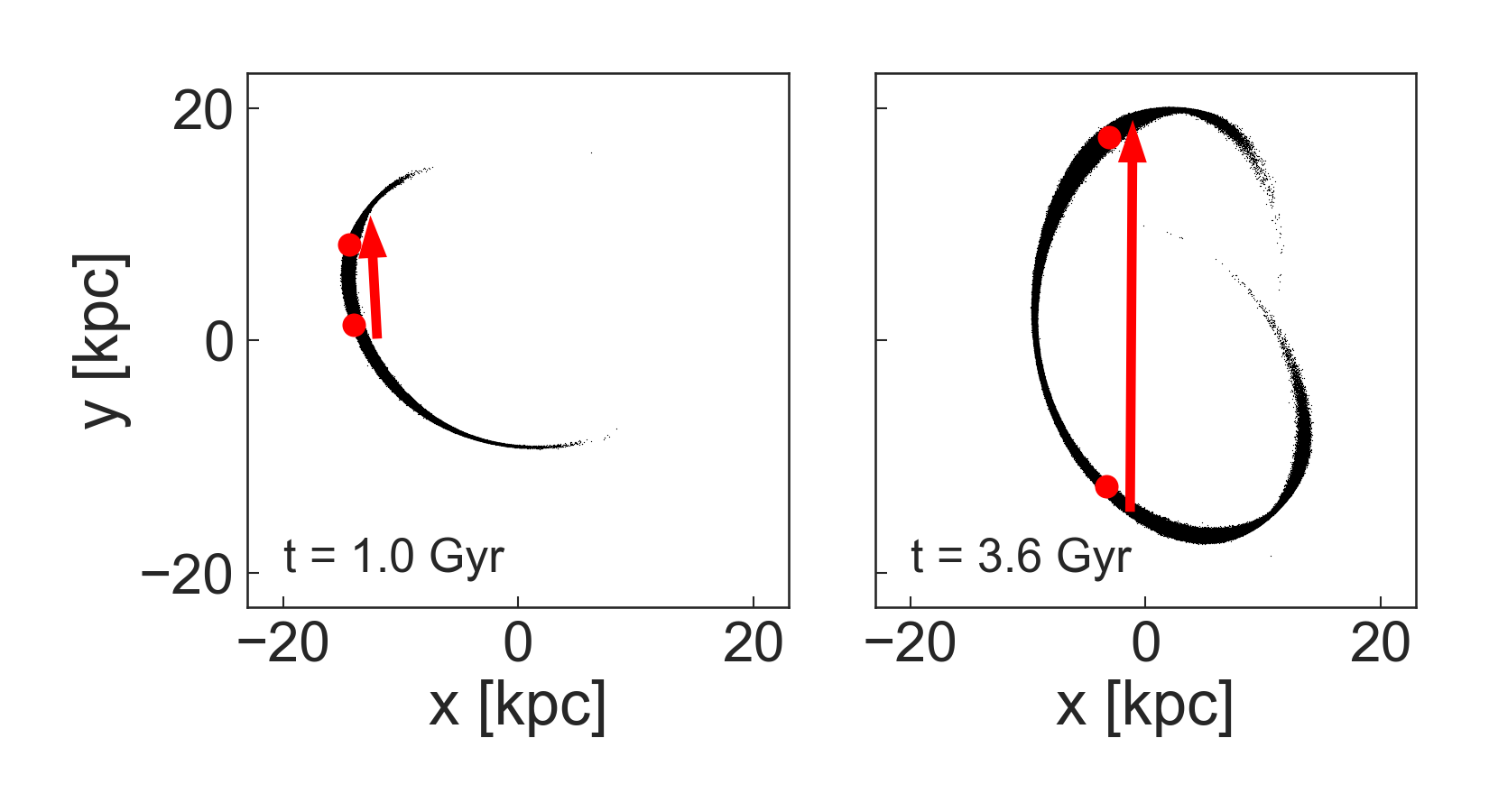}
  \caption{Left: A stream $1$ Gyr after an interaction with a subhalo. The two red dots indicate the orbits that are being used to measure the size of the gap, the red arrow shows the distance on a straight line between the two dots. Right: The same stream, $3.6$ Gyr after the encounter. At this point in time the two orbits (red dots) are close to their maximum separation.}
  \label{fig:maxsizeplot}
\end{figure}

We compare the predictions of the model presented in
Sec.~\ref{sec:AA2orb} and Sec.~\ref{sec:centraldensity} with the gaps
produced in the N-body experiments. We will first investigate gaps produced by
subhalos of varying mass and size for a fixed encounter
configuration (i.e. the same velocity and impact angle). Next, we focus on
the effects of a varying configuration while keeping the subhalo~properties~fixed.

\subsection{Size evolution}

Figure~\ref{fig:modelvsnbody} shows the evolution of gaps caused by
interactions sharing the same configuration, but with different
subhalo masses (see Table~\ref{table:subhalos} for their
properties). The model (solid lines) reproduces very well the size of
the gap as measured in the N-body simulation (coloured dashed
lines). The latter is measured as the average separation of two groups
of 50 particles on each side of the gap. These 50 particles are
identified as those that experience the largest velocity change at the
time of the impact. We use $50$ particles to lower the effects of
discreteness of the N-body simulation, but there is only very little
difference when using the single particle with the maximum velocity
change on each side of the gap. The bottom panel of
Figure~\ref{fig:modelvsnbody} shows the total distance of the gap to
the centre of the host galaxy and gives an indication of its
orbit. The frequency of $r(t)$ and the oscillations in the gap size
are in antiphase. This is naturally expected since the gap will be
stretched at pericentre and be smallest at apocentre.

Although Fig.~\ref{fig:modelvsnbody} shows that the model reproduces the gap size in the N-body experiment very well, there appears to be an upper limit to its measured size. The largest difference is apparent for the encounter with the most massive subhalo at late times. This limit occurs because the size of the gap becomes comparable to the typical scale of the orbit and hence our method of measuring the size of the gap fails to work. The typical scale of the specific orbit that is used here is $\lesssim 40$~kpc, see Fig.~\ref{fig:maxsizeplot}. This value can also be determined analytically using the inverse of Eq.~\eqref{eq:xyz2AA}, and considering that the two orbits on each side of the gap are at a maximum separation at $\Delta\boldsymbol{\Theta} = \pi$. The maximum distance between two particles on the same orbit but apart by $180^\circ$ in the angles, at any location in the orbit, is $\sim 35$ kpc. This value agrees very well with the ceiling reached by the red, dashed line measured from the N-body experiment in Fig.~\ref{fig:modelvsnbody}.

The size of the gap in this regime is pushing the limits of our analytical model. The transformation from action-angles to Cartesian coordinates (i.e. Eq.~\ref{eq:xyz02xyzt}) is only valid `locally' near the central orbit, and therefore the approximation breaks down for such large gaps. Although it should be possible to extend the formalism to include such cases, this is not really necessary as there are no known streams with gaps of this size - nor is it likely to observe one such gap in the (near) future. 

\subsection{Evolution of the density}\label{sec:evolofdens}
\begin{figure*}
	\centering
    \includegraphics[width=0.9\hsize]{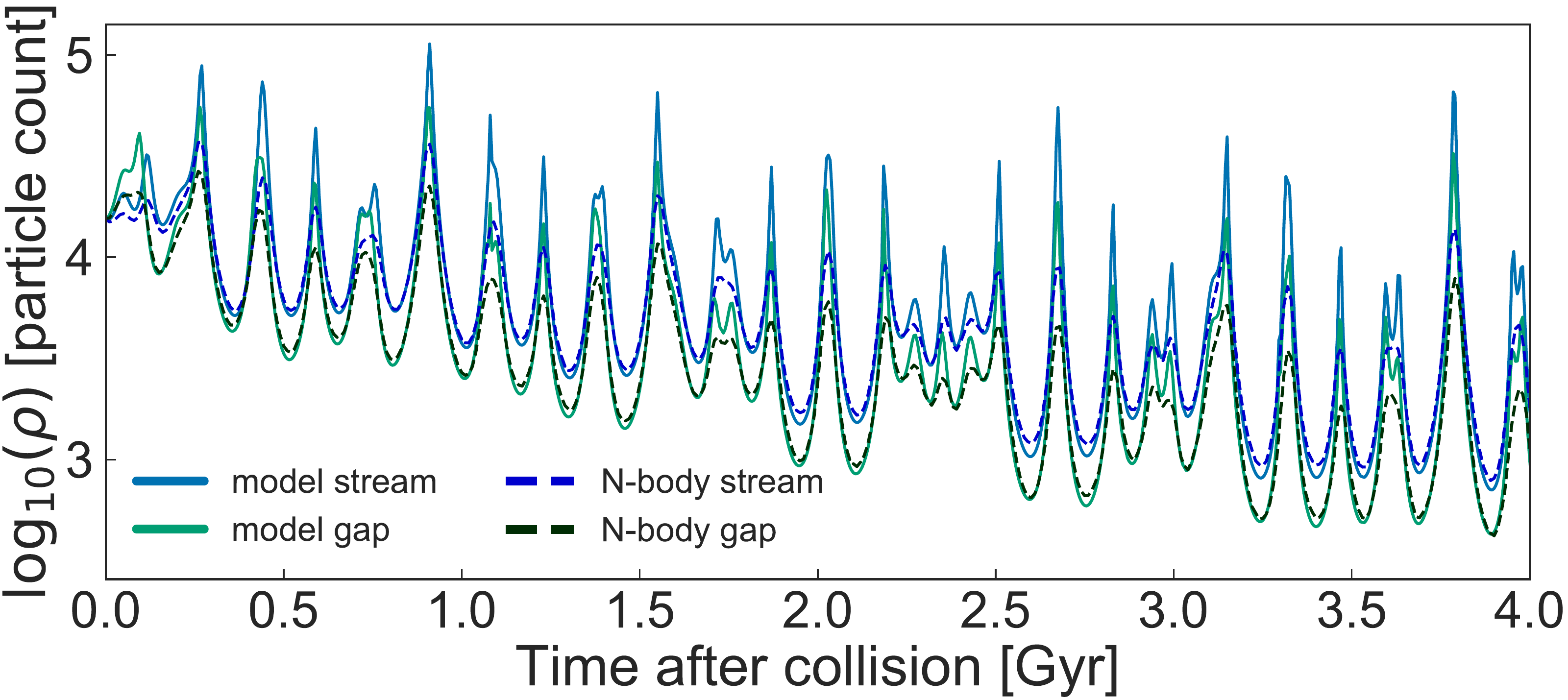}
    \caption{Density evolution of an unperturbed stream (in blue) and
      of a gap in a stream following the same orbit (in green). The
      dashed curves are the densities obtained from the corresponding
      N-body experiments. The agreement between the solid and dashed
      curves is excellent. The subhalo that is used to create the gap
      in the stream has a mass of $10^7\,\mathrm{M_\odot}$.}
    \label{fig:density_gap_vs_nogap}
\end{figure*}
Now we compare the density as predicted by our model with the density measured in N-body experiments. For the latter we count the number of particles in a small volume in 6D space with $r<0.1$ kpc and $v < 7.5$ km/s. This velocity limit does not remove any particles from the stream at $t_0$ when the impact occurs, but it removes particles that may have drifted away (i.e. have a different orbital phase) at later times. The volume is centred on the central orbit, which is determined in a simulation of a stream with the same set-up, but without a subhalo interaction.

Figure~\ref{fig:density_gap_vs_nogap} compares the predicted (solid
lines) and measured density from the N-body simulation (dashed lines)
for a stream with and without a gap. For the latter, we have simulated
the exact same stream with and without an encounter with a
subhalo. Although the peaks and troughs of the stream and the gap are
always larger in our model, the figure shows that the model provides
an excellent description of the N-body experiments. The small
differences can be attributed to a resolution effect: in the N-body
experiments we measure the density in a finite volume, whereas the
model computes a density at a single location in space. If the density
were measured in a smaller volume in our experiments the peaks would be
sharper. However, the number of particles would drastically decrease
and drop to less than a handful in less than $4$~Gyr of evolution.

\subsubsection{Varying subhalo masses}

\begin{figure}
	\centering
    \includegraphics[width=\hsize]{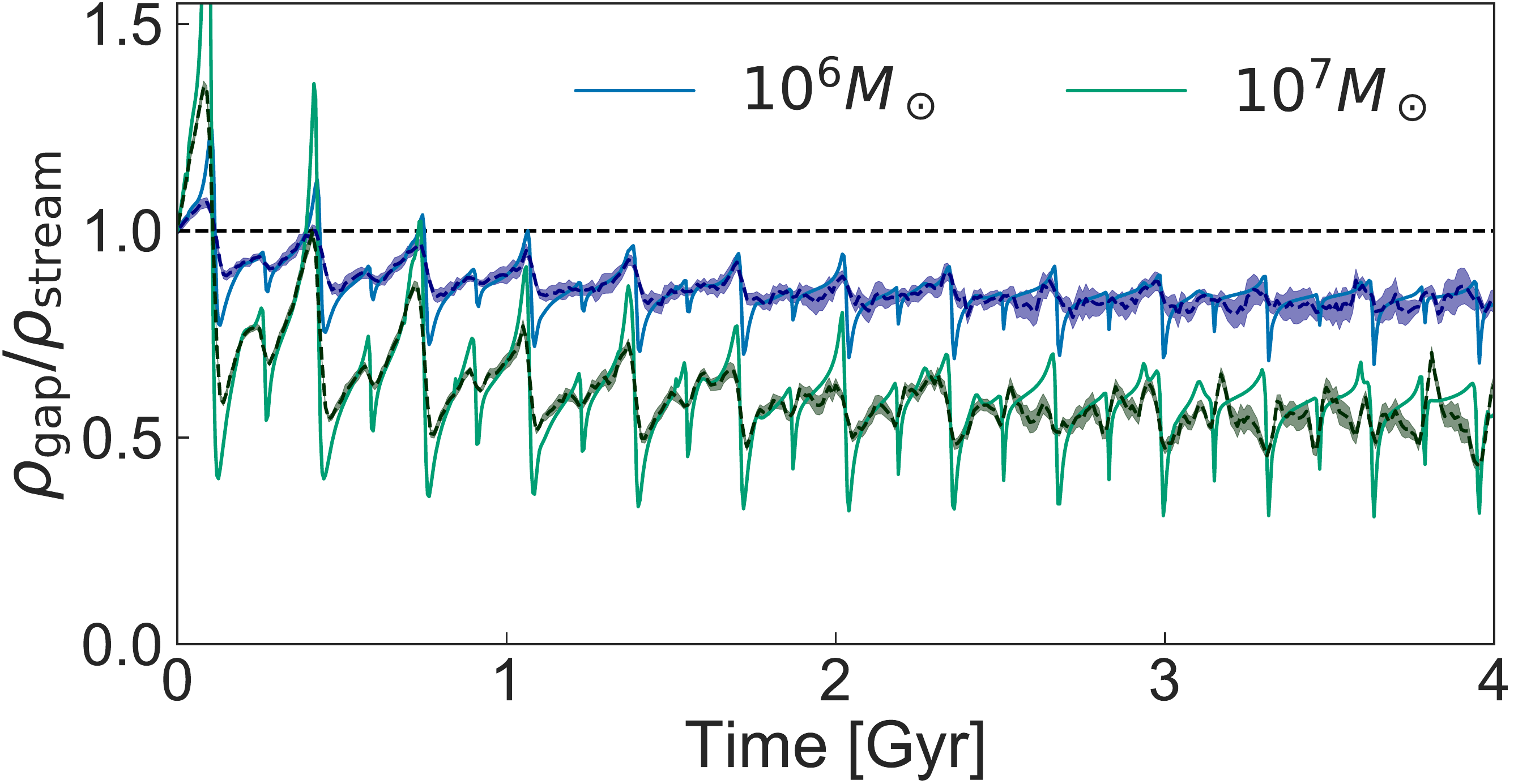}
    \includegraphics[width=\hsize]{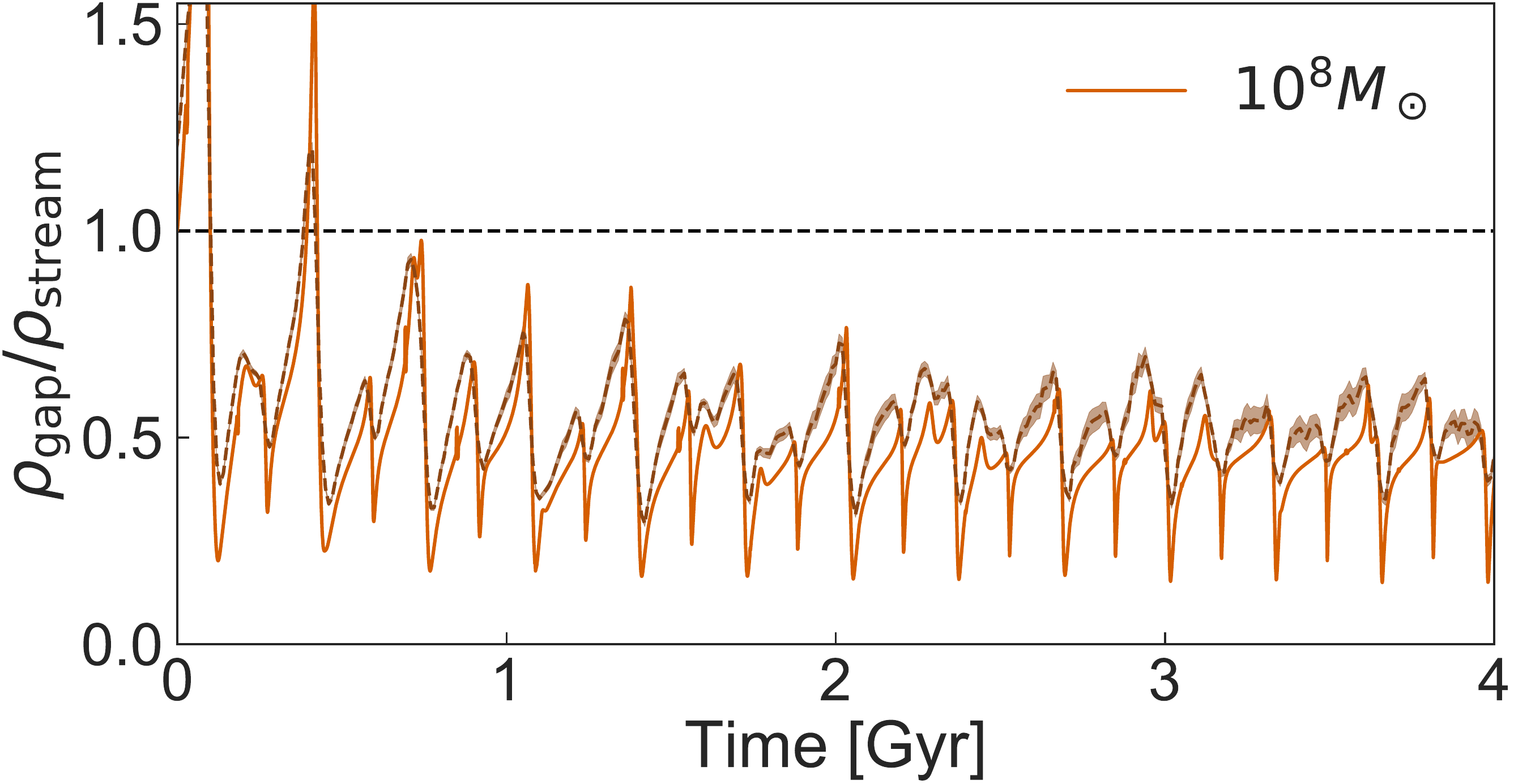}
    \caption{Density contrast measured for a stream experiencing an encounter with subhalos of different mass and scale radius, see Table~\ref{table:subhalos}. The relative velocity of the impact is the same for all three interactions. The coloured solid lines show the density contrast according to our model, and the dashed lines are measured in the N-body experiments. The shaded areas indicate the Poisson error in the observed density. The subhalo of $10^8~{\rm M}_\odot$ is shown in a different panel because of its slightly different set-up.}
    \label{fig:density_difference_multiplemasses}
\end{figure}

Next, we explore the evolution of the density contrast (i.e. the ratio
of the density around the gap to that of the unperturbed stream) in
Fig.~\ref{fig:density_difference_multiplemasses}. The figure shows the same
experiments as those plotted in Fig.~\ref{fig:modelvsnbody}, with the
density contrast of the most massive subhalo shown in a separate
panel. For the most massive halo, we have modified slightly our set-up, instead of starting from the same initial conditions as the other
experiments using the orbit shown in Fig.~\ref{fig:orbit}, we have
used the location of the gap to determine its orbit. We used
this as the central orbit both in our analytical model and for the
N-body experiment representing the unperturbed stream. The reason for
this is that when the subhalo and the stream interact, the stream
receives an impulse that displaces it slightly from its original
orbit. The effect is negligible for subhalos of
$M \lesssim 10^7 M_\odot$, and is small but apparent for more massive
objects, particularly after $\sim 3 - 4$~Gyr of evolution. This new set-up is
actually more realistic since when attempting to model an observed
stream or gap, its actual measured position and velocity in a suitable
gravitational potential would be integrated (as it is not possible to
have a priori access to the original initial conditions of the orbit
of the stream, before it received the impact).

In Fig.~\ref{fig:density_difference_multiplemasses}, we show with
solid lines the predicted density contrast from our model, and with dashed
lines those measured in the N-body experiments. The Poisson
errors on the ratio of the densities as measured in the N-body
experiments are marked with shaded areas.
In general, the amplitude of the density contrast is well reproduced
by the model, with 
the difference in the amplitude of the narrow peaks at
early times explained by the same resolution effects as described in 
Sec.~\ref{sec:evolofdens}.

\subsubsection{Variation of encounter configuration}

Finally, we check how our model performs for different configurations
of the stream-subhalo encounter, keeping the subhalo at a fixed mass
of $10^7~{\rm M}_\odot$. We compare three different configurations which
correspond to rotations of the same velocity vector, as listed in
Table~\ref{tabel:configurations}, with Configuration 1 being that used in the previous section. We note that the velocity vector is
rotated in the rest frame, not in the co-moving frame of the
stream. The resulting configurations thus have different velocity
amplitudes in the co-moving frame. 

\begin{figure}
	\centering 
    \includegraphics[width=0.975\hsize]{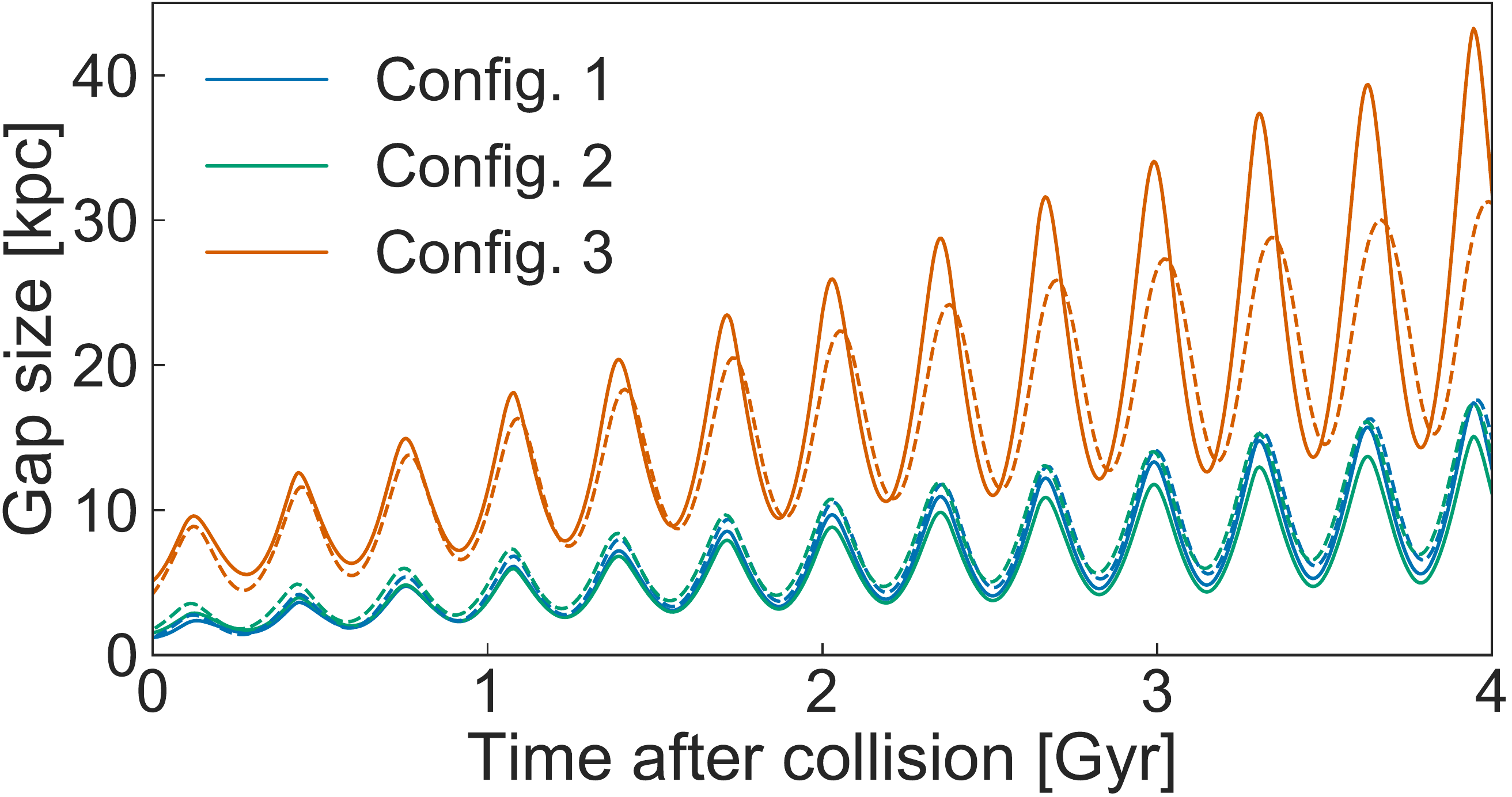}
    \caption{Similar to Fig.~\ref{fig:modelvsnbody} but for three different configurations of the stream-subhalo interactions. The gaps of all configurations are created by the same subhalo of size $10^7\mathrm{M_\odot}$, but their relative velocity $w$ and impact angles $\theta$ and $\alpha$ are different.}
    \label{fig:sizes_multipleconfigurations}
\end{figure}

\begin{figure}
	\centering
    \includegraphics[width=\hsize]{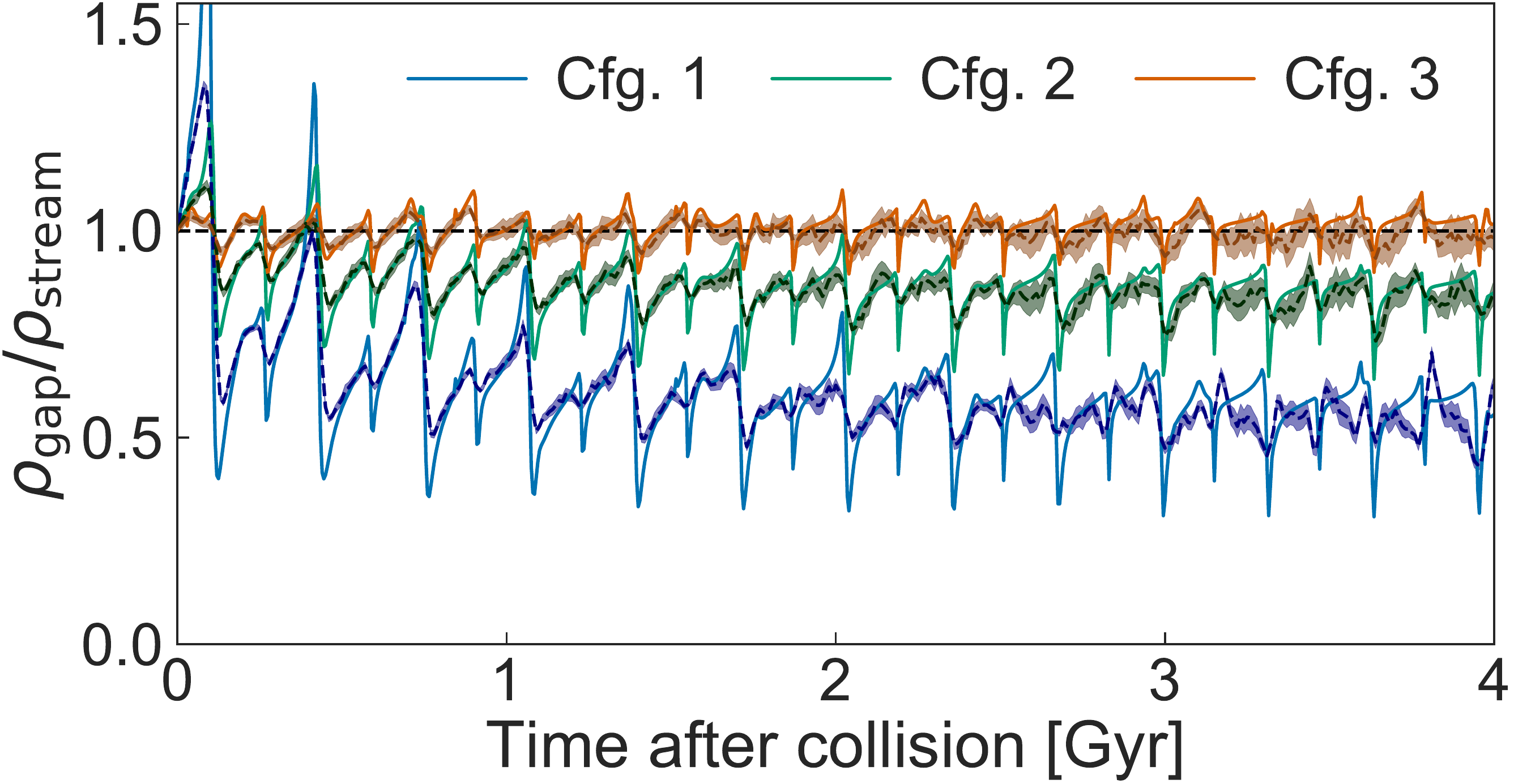}
    \caption{Similar to Fig.~\ref{fig:density_difference_multiplemasses}, but for three different stream-subhalo interactions with a subhalo of fixed mass, and whose size evolution is described in Fig.~\ref{fig:sizes_multipleconfigurations}.}
    \label{fig:density_difference_multipleconfigurations}
\end{figure}

Figures~\ref{fig:sizes_multipleconfigurations}~and~\ref{fig:density_difference_multipleconfigurations} show the time evolution of the size and density contrast. The layout of the figures is the same as in the previous section. Again, the model (solid lines) predicts the behaviour of the gaps as measured in the N-body experiments (dashed lines) extremely well. 

Interestingly, Configuration 2 with a subhalo of
$10^7~\mathrm{M_\odot}$ gives rise to a density contrast of similar
amplitude as the collision with the subhalo of $10^6~\mathrm{M_\odot}$
(see Fig.~\ref{fig:density_difference_multiplemasses}) on Configuration 1, both producing
a gap with a density contrast of $\sim 0.9$. However, if we examine the size of the gaps we notice that the gap
caused by the $10^6~{\rm M}_\odot$ subhalo (blue curve
Fig.~\ref{fig:modelvsnbody}) is smaller than that for the $10^7~{\rm M}_\odot$. This implies that by measuring both size and density, one may be sensitive to different parameters characterising the encounter, as we shall discuss in more detail in Sec.~\ref{sec:observables}.

\begin{figure}
	\centering
    \includegraphics[width=\hsize]{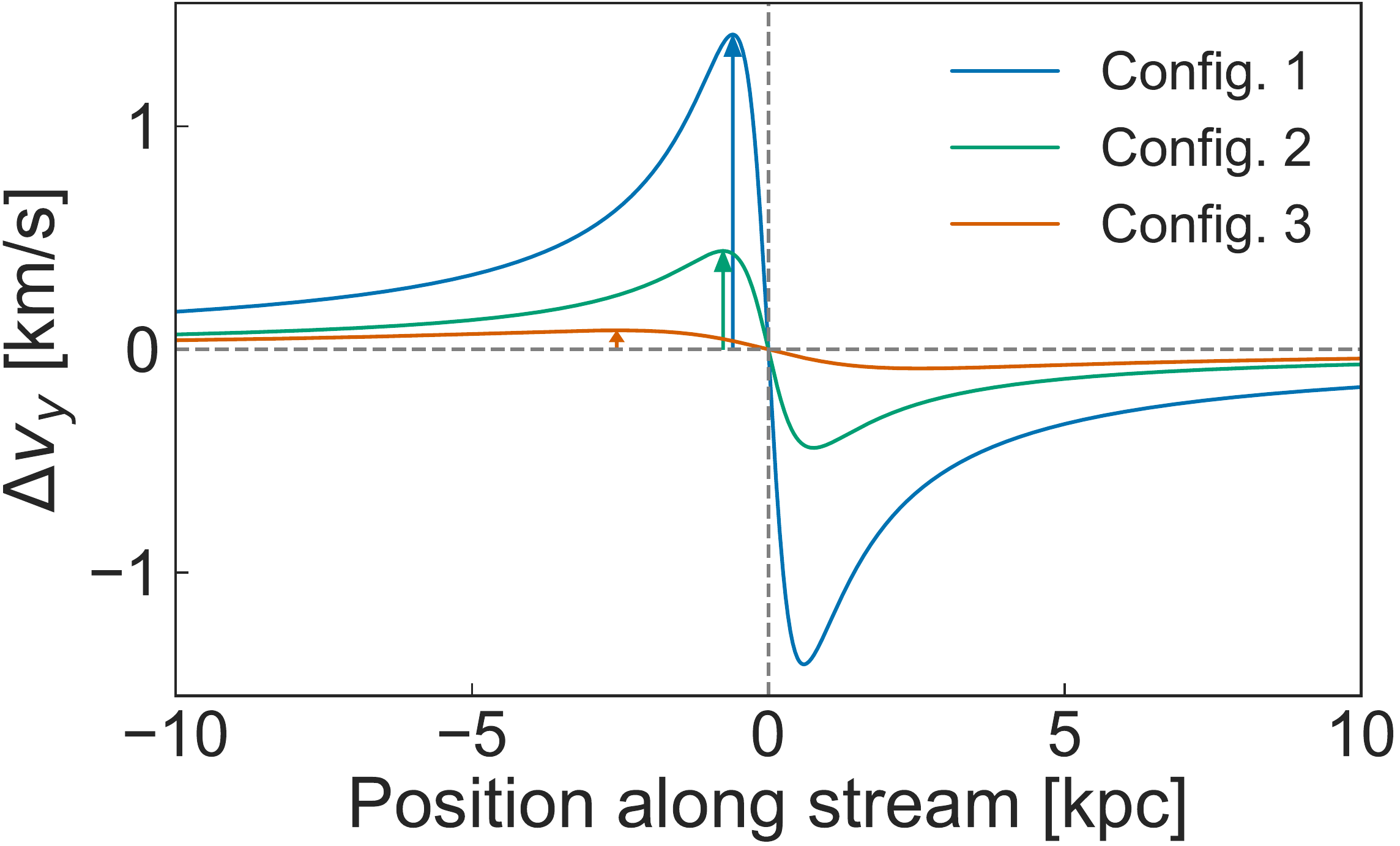}
    \caption{Profile of the change in velocity along the stream $v_y$, for the three different configurations described in the two previous figures. In Configuration 1 the stream experiences the largest kick on the smallest scale. On the other hand, in Configuration 3 the kick is much smaller but on a much larger scale. The arrows indicate the location of the maximum $\Delta v_y$.}
    \label{fig:kickprofile_configurations}
\end{figure}

Figure~\ref{fig:sizes_multipleconfigurations} shows that the gaps
resulting from the encounter in Configurations 1 and 2 have initially
approximately the same size. Interestingly the gap resulting in
Configuration 3 is initially the largest and also remains the largest
throughout its evolution in time (although the size is somewhat poorly
modelled because of the change in velocities for this particular
configuration is very shallow, see
Fig.~\ref{fig:kickprofile_configurations}, which gives rise to
some complications in identifying the correct particles to trace in the
N-body).

On the other hand, Fig.~\ref{fig:density_difference_multipleconfigurations} shows that 
the evolution of the density for Configuration 3 is very similar to that of an unperturbed stream. To understand this consider the velocity kicks $\Delta v_y$ and their profiles as shown in Fig.~\ref{fig:kickprofile_configurations}. Since the initial size of the gap is derived from the location of the maximum velocity change, the gap produced in Configuration 3 initially is much larger than the other two. Comparing Fig.~\ref{fig:density_difference_multipleconfigurations} and Fig.~\ref{fig:kickprofile_configurations} we see that the steepest density contrast is associated with the largest change in velocity, exactly like one would expect. \emph{These results imply that the size of the gap is strongly correlated with the distance between the maximum velocity change, whereas the density contrast is more correlated with the amplitude of the change} (see also the expressions in the next section).

\begin{table}
\caption{Parameters of the different configurations, see Sec.~\ref{sec:impulse} for their definition.}
\label{tabel:configurations}
\centering
\begin{tabular}{c c c c c}
\hline\hline
 & Config. 1 & Config. 2 & Config. 3 \\
\hline      
$\theta$ [deg.] & -10.5 & -34.1 & 97.6 \\
$\alpha$ [deg.] & 39.4 & 9.9 & 245.2\\
$w$ [km/s] & 76.7 & 33.7 & 378.3 \\
\hline
\end{tabular}
\end{table}

\section{Exploration of the gap observables: dependencies and degeneracies}\label{sec:observables}

Now that we have validated our analytic model, we will use it to explore how the size and density of a gap depend on the collision parameters using Eqs.~\eqref{eq:Dx_pred} and \eqref{eq:density_func_of_time}. We will consider hypothetical gaps formed in the stream presented in Sec.~\ref{sec:Nbody} and analysed in Sec.~\ref{sec:results}. To this end, we vary the characteristic parameters of the collision with a subhalo, namely $w$, $\theta$, and $M$, while keeping the angle $\alpha$ fixed at some arbitrary value $\alpha = 163^\circ$. We consider $w$ in the range $[0,800]$~km/s and $\theta$ in the range $[-90^{\circ},90^{\circ}]$. Instead of varying separately $M$ and $r_s$, we use a relation for $V_\mathrm{max} \propto r_\mathrm{max}$ for subhalos found in the `Aquarius' simulations by \cite{Springel2008}, see Appendix~\ref{sec:subscarel} for details. 

\begin{figure*}
	\centering
    \includegraphics[width=\hsize]{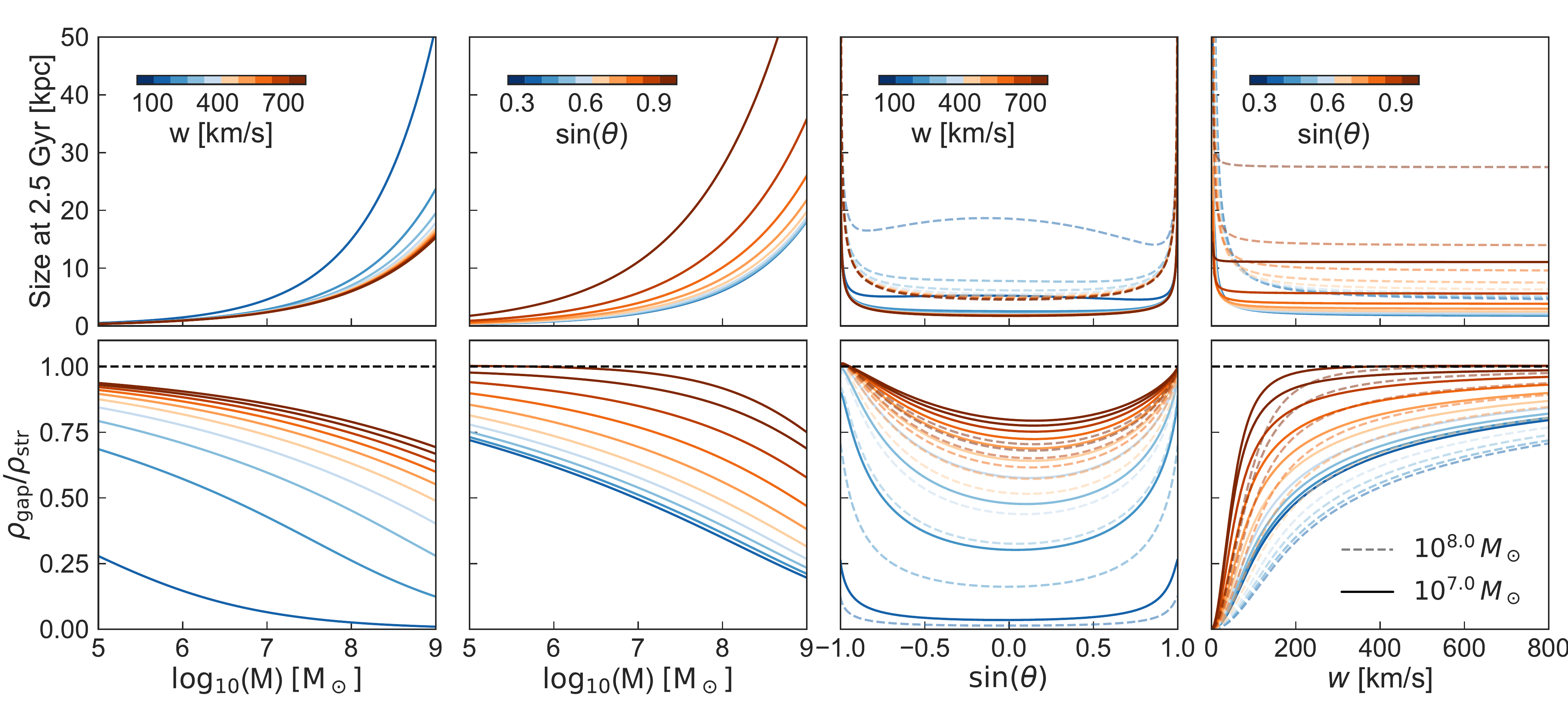}
    \caption{Top row: size of the gap after 2.5 Gyr of evolution, as a function of the impact parameters (subhalo mass, angle $\theta$ and amplitude of the relative velocity $w$). Bottom row: density contrast as a function of the same parameters. The two panels on the left show the discrete variation of the curves - varying the velocity amplitude $w$ or collision angle $\theta$. The two panels on the right show two sets of lines, solid for a subhalo of mass $10^{7}\,\mathrm{M_\odot}$ and dashed for $10^{8}\,\mathrm{M_\odot}$.}
    \label{fig:size_densitycontrastboth}
\end{figure*}

Figure~\ref{fig:size_densitycontrastboth} shows the gap's properties, namely size and
density contrast, 
as a function of these characteristic parameters.
Each panel shows the dependence of these two observable quantities
with one of the three parameters: $M$, $w$, or $\sin{\theta}$. At the same time,
we vary discretely a second parameter, which gives rise to the
different curves in each subpanel, but keep fixed the third parameter. For example, in
the leftmost panels we show the variation of gap size (top) and
density contrast (bottom) 2.5~Gyr after impact as a function of mass
of the subhalo $M$, for different values of $w$ as indicated by the
colour bar, and for $\theta = \pi/4$.

Since the size of a gap varies depending on its orbital phase, we have
checked that the dependencies shown in
Figure~\ref{fig:size_densitycontrastboth} are robust. We have found
them to be identical, except for an overall scaling of the amplitude that
depends on the phase. Since it will be possible to establish the
phase of a gap observationally (after assuming a suitable Galactic
potential and integrating the orbit of the stream in which it is
embedded), this implies that the trends observed here can be used to
infer several of the characteristic properties of the encounter. The
density contrast, meanwhile, does not vary along the orbit (because
the density variations along the orbit for the gap are identical to
those for the unperturbed stream). However, in the bottom panels of Figure~\ref{fig:size_densitycontrastboth}, we have
taken the late times limit of the density contrast given by Eq.~\eqref{eq:dens_plateau}.

We have shown in Eq.~\eqref{eq:maxkickpos} that the size of the gap at
any point in time strongly depends on its initial magnitude (i.e.
$\propto r_s/\cos{\theta} $) implying a dependence on the subhalo's
mass through the $r_s$ parameter (with $r_s \propto M^{2/5}$), as shown in 
Appendix~\ref{sec:subscarel}. This simple relation
explains the curves in the top panels of
Fig.~\ref{fig:size_densitycontrastboth} well, which show that the gap
size depends strongly on the mass of the subhalo (two leftmost
panels), with relatively little dependence on $w$ and $\sin\theta$, except
for extreme values of these parameters (two rightmost panels). For
example, when the subhalo moves along the stream (i.e. when
$\cos{\theta} \to 0$) the size of the gap is clearly not well
defined. In this case the impulse approximation breaks down as the
subhalo and stream interact for a long time, and, perhaps more
importantly, the interaction affects a large part of the stream. Also
for very low values of $w$, the impulse approximation is no longer
valid. Low relative velocities and extreme alignment must be
rare because they only happen when the stream and subhalo move at a
similar velocity and in the same direction. In summary, the top panels of
Figure~\ref{fig:size_densitycontrastboth} suggest that given a
gap size, it is possible to infer with some confidence the mass of the
subhalo that perturbed it for most values of $w$ and $\theta$.

With knowledge of the mass, the density contrast could be used to
infer some plausible encounter geometries. To understand the factors
driving the density contrast, we use Eq.~\eqref{eq:dens_plateau},
which depends on the ratio of $\det |\sigma_{\Theta_0}|$ for the
stream and the gap. Although general analytic expressions can be
obtained, these are somewhat cumbersome. Under the assumption that the
velocity kick is small (compared to the characteristic orbital
velocity of the subhalo), and that the covariance matrix of the stream
at the time of the encounter is diagonal in Cartesian coordinates, we
find (see Appendix~\ref{Ap:expl_dep} for full details)
\begin{equation}
\delta\rho_\mathrm{str}^\mathrm{gap} \propto 1 - \frac{GM}{r_s^2w} f(\theta, \alpha, \mathbb{C}^{\rm
  str}_{{\boldsymbol{x},\boldsymbol{x}}_0}, \mathbb{C}^{\rm
  str}_{{\boldsymbol{v},\boldsymbol{v}}_0},
\boldsymbol{x}_0,\boldsymbol{v}_0), 
\label{eq:drho_analytic}
\end{equation} 
where
$f(\theta, \alpha, \mathbb{C}^{\rm
  str}_{{\boldsymbol{x},\boldsymbol{x}}_0}, \mathbb{C}^{\rm
  str}_{{\boldsymbol{v},\boldsymbol{v}}_0},
\boldsymbol{x}_0,\boldsymbol{v}_0)$ is a function that depends on the
angles characterising the encounter, the location of the encounter
$(\boldsymbol{x}_0,\boldsymbol{v}_0)$ and the configuration and
velocity covariance matrices of the unperturbed stream at the time of
the impact (see also Eq.~\ref{eqapp_rho}). This relation implies that
the density contrast becomes shallower with increasing $w$, as can
indeed be seen in Fig.~\ref{fig:size_densitycontrastboth}. On the
other hand, more massive subhalos create gaps with lower
densities. Because of the different dependence of the gap size (top
panel) and of the density contrast with the characteristic parameters
of the encounter, $(M, w, \sin{\theta})$, this means that it is
possible to break some of the degeneracies present using these two
observable quantities, provided the time since the collision could be
established (which is necessary for making use of the constraints
provided by gap~size).

\begin{figure*}
  \includegraphics[width=\hsize]{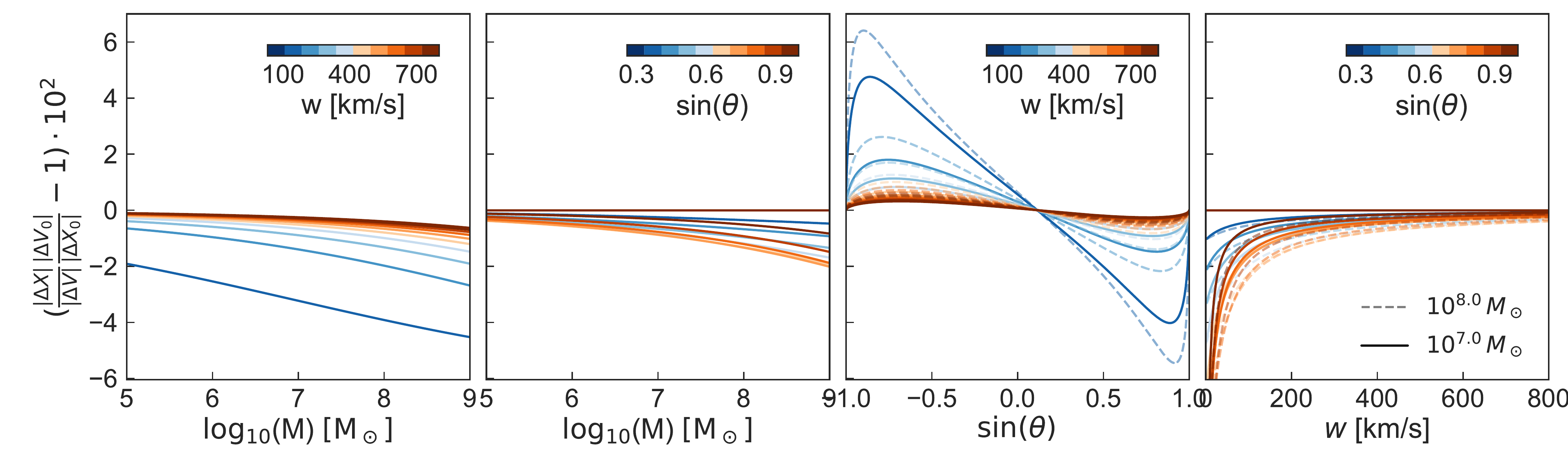}
  \caption{Sensitivity of the (time-independent) ratio of the size of the gap to the separation in velocity to the same parameters discussed in Fig.~\ref{fig:size_densitycontrastboth}.}
  \label{fig:dsize}
\end{figure*}
Another time-invariant combination of observables is plotted in
Fig.~\ref{fig:dsize}. This figure shows the spatial size of the gap
relative to the separation in velocity space, normalised by its
initial value at $t_0$. Note that although the ratio of $\Delta X$ and
$\Delta V$ does vary with the orbital phase of the gap/stream, 
this phase can be established through orbit integrations as discussed
earlier. The lines shown in Fig.~\ref{fig:dsize} (measured at 2.5
Gyr) are for a gap that is near its apocentre. Evaluating the ratios
$\Delta X/\Delta V$ near the pericentre results in a similar figure,
but where the $y$-axis is mirrored with respect to the line $y=0$.

There are clear similarities between this ratio and the behaviour of
the density contrast shown in the bottom row of
Fig.~\ref{fig:size_densitycontrastboth}, except for the third panel,
which reveals a sensitivity for low $w$ velocity on the angle of the
encounter $\theta$. Overall, this ratio will therefore be a good discriminator for 
low $w$ of $\sin \theta$.

In summary (smallish) gaps $\lesssim 25$~kpc are mostly dependent on
the mass of the subhalo, while large gaps can either be due to a
specific configuration (low relative velocity or angle of the
encounter) or due to a large subhalo. Assuming the average encounter
has a relative speed of $w>200$~km/s it appears that per orbit/stream
we can break the degeneracy of the interaction parameters using also
the density contrast. We have checked that these conclusions (and
dependencies) are robust and independent of the orbital
characteristics of the stream (e.g. different inclinations with
respect to the Galactic plane), but they are only strictly valid if
the age of the gap can be well constrained.

\section{Discussion}\label{sec:CandD}

Action-angle variables have been previously used to describe streams
and their gaps \cite[e.g.][]{Helmi1999, Helmi2007, Bovy2014,
  Sanders2016DynamicsInteractions, Helmi2016,
  Bovy2017LinearSpectrum}. There is a trade-off to be made when using
these variables: one may either make use of a numerical approach and
obtain a (local) approximation for a generic potential \citep{Binney2012, Sanders2015}, or use a
fully analytic approach and be restricted in the choice of the
potential. In this work, we take the latter approach such that we
can express the properties of the gaps in physical space directly as a
function of the~encounter~parameters.

In contrast to the work of \cite{Erkal2015}, who argue that gaps grow
at late times as $\sqrt{t}$ (for circular orbits), we find that both
in our numerical experiments as in the analytic model, gaps grow
linearly with time independently of the type of orbit or shape of the
gravitational potential. We have thus extended the findings of HK16
who considered a spherical potential, and confirm also the results of
\mbox{\cite{Sanders2016DynamicsInteractions}}.

\cite{Sanders2016DynamicsInteractions} have found that the density
contrast of a gap approaches a constant value at late times. Our fully analytical
model allows us to verify their conclusion and we are also able to
show why this happens and what the constant value depends upon
(e.g. Eqs.~\ref{eq:dens_plateau} and
\ref{eq:drho_analytic}). \cite{Sanders2016DynamicsInteractions} also
find that gaps grow differently in the leading and trailing
arm. Judging from the expressions derived in this work, there may
be two reasons for the different growth rate: $i$) a difference in the
local (velocity) dispersions of the particles in either the leading or
trailing arm, as was already noted by
\cite{Sanders2016DynamicsInteractions}; $ii$) the orbits of the
leading and trailing stream have slightly different characteristic parameters (they
are slightly offset in energy), and these affect the growth rate of
gaps as well as the decline in their density.

\cite{Erkal2015b} argue that there exists a degeneracy in the gap
parameters with mass and velocity. The reported degeneracy of
$(M,{\bf w})\rightarrow(\lambda M,\lambda{\bf w})$ only exists if the
scale radius $r_s$ of the subhalo is kept fixed, but its mass is not.
For example, the size of a gap depends on $r_s$, while the
density contrast depends on $M/r_s^2$. Since a
non-linear relation between $r_s$ and $M$ is known to exist for
subhalos in cosmological simulations ($r_s \sim M^{2/5}$, \citealt{Neto2008TheConcentrations, Springel2008}), we must conclude that
the above degeneracy does not exist.

The model presented in this work successfully describes gaps in streams in axisymmetric potentials. However, it builds on several key assumptions, namely:

\begin{itemize}[noitemsep,topsep=0pt]
 \item {\it The time of the collision/age of the gap}. Although this information is in principle encoded in the size of the gap, we have generally explored its properties at a fixed time. Keeping it open will add one more parameter to optimise. A rough estimate of the formation time could be obtained by integrating the two sides of the gap backwards in time and to see when they meet. Recall on the other hand, that if we may assume that the encounter happened sufficiently long ago, the density contrast is time- independent, and some of the encounter parameters can be constrained.
 
 \item {\it The potential of the host galaxy}. Changes in the host potential will change the central orbit of the gap and thus the size and density evolution. However, we note that the explicit time dependence of both the size and density of the gap will not change with (small) variations in~the~potential. 

 \item {\it Knowledge of the pristine stream conditions, before the interaction}. In principle from observing the full stream morphology and knowing the age of the gap, it should be possible to derive the full 6D properties of the stream at the time of~the~collision.

 \item {\it The properties of the subhalo}, here assumed to be well-described by a Plummer sphere. This choice was made because of its simple mathematical expression, for which there is an analytic solution to the integral of the impulse approximation used
to compute the velocity kicks. However, it is possible to compute this integral numerically for other profiles \citep[e.g.][]{Sanders2016DynamicsInteractions}.
\end{itemize}
\vspace{0.5cm}

We have shown here that some of the previously reported degeneracies in the space of parameters describing the encounter can be broken by making plausible assumptions on the stream-subhalo configuration. A natural next step would be to consider probability distributions for the encounter parameters, much like, for example \cite{Erkal2016a}. Our model can then be used to quickly explore the parameter space as it can constrain the most likely encounter parameters given observation of~a~gap. 

\section{Conclusions}
We have successfully extended the model of the evolution of gaps in spherical potentials, presented in HK16, to describe gaps in streams orbiting in axisymmetric St\"ackel potentials. The model accurately predicts the evolution of both the size of the gap and its central density. The model is unique in that it is \emph{fully analytic}, meaning that we can directly relate the stream-subhalo interaction parameters to the properties of the resulting gaps. In doing so, it provides some interesting insights into the evolution of gaps in streams.

We find that the sizes of the gaps in axisymmetric potentials grow
linearly in time~-~and this dependence is independent of the shape of
the Galactic potential. On the other hand, the density declines in
time as $t^{-n}$ where $n$ denotes the number of independent
frequencies characterising its orbit. The growth of the size and
density of a gap depend on the subhalo properties (mass and scale
radius), the properties of the stream at the time of the impact
(velocity and positional differences of the particles), and on the
central orbit of the gap.

We have shown that the size of the gap is correlated with the portion
of the stream most affected by the subhalo flyby (the value
$y_{\rm max}$ in the impulse approximation). The density contrast of
the gap, on the other hand, is more correlated with the amplitude of
the interaction ($\Delta v^{\rm max}$). These different correlations
are in the end, what drives the ability to break the degeneracy of the
encounter parameters. For example, for a given gap age, small gaps
($<25$~kpc) are very dependent on the size of the subhalo, while a
large gap can be caused by a large subhalo, or by an alignment of the
orbit of the stream and subhalo. These results are encouraging and
appear to be useful to constrain the properties of a population of
dark subhalos if present in the halo of the Milky Way.

\begin{acknowledgements}
We gratefully acknowledge financial support from a VICI grant and a Spinoza Prize from the Netherlands Organisation for Scientific Research (NWO) and HHK is grateful for the support from the Martin A. and Helen Chooljian Membership at the Institute for Advanced Study. We thank Tim de Zeeuw for his comments on an early version of this manuscript and Hans Buist for modifications made to the code used here. For the analysis, the following software packages have been used: {\tt vaex} \citep{Breddels2018}, {\tt numpy} \citep{VanDerWalt2011TheComputation}, {\tt matplotlib} \citep{Hunter2007Matplotlib:Environment}
, {\tt jupyter notebooks} \citep{Kluyver2016JupyterWorkflows}, {\tt f2py} \citep{Peterson2009F2PY:Programs}, and {\tt pyGadgetReader} \citep{Thompson2014PyGadgetReader:Python}.
\end{acknowledgements}

\bibliographystyle{aa} 
\bibliography{references}

\appendix
\onecolumn
\appendix

\section{Covariance matrix - full 3D impulse}\label{sec:3D_cov}
In this section we derive the change of the covariance matrix when the full 3D morphology of the stream is taken into account. Similar to the 1D case, see Sec.~\ref{sec:centraldensity}, we assume that the change in velocity is a linear function of the spatial coordinates, meaning the denominator of the kicks $(w( (r_s^2 + r^2)w^2 - (\vec{x}\cdot\vec{w})^2)) \approx r_s^2w^3$. This approximation is in general true for the small volumes in which we measure the density, typically $<< 1$ kpc. This assumption allows us to rewrite Eqs.~\eqref{eq:3Dv_equations} to
\begin{equation}
 \Delta v_i(\vec{x}) = -2GM\frac{w^2x_i - w_i(\vec{x}\cdot\vec{w})}{r_s^2w^3},
\end{equation}
where the subscript ${i} = x,y,z$.

In a similar procedure as for the 1D approximation we can now compute the covariance terms. The velocity-position terms in the most general form are
\begin{equation}
C(v_i + \Delta v_i, x_j) = C(v_i,x_j) + \frac{1}{n}\sum\Delta v_i(x_j-\bar{x}_j)
\end{equation}
with
\begin{equation}
\frac{1}{n} \sum{\Delta v_i (x_j -\bar{x}_j)} = -\frac{2GM}{r_s^2w^3} \bigg[w^2C(x_i,x_j) - \sum_{k=x,y,z} w_iw_kC(x_k,x_j)\bigg].
\label{eq:apA1}
\end{equation}
The velocity-velocity terms are a bit more cumbersome
\begin{equation}
C(v_i + \Delta v_i, v_j+\Delta v_j) = C(v_i,v_j) + \frac{1}{n}\sum\Delta v_i(v_j-\bar{v}_j)
+ \frac{1}{n}\sum(v_i-\bar{v}_i)\Delta v_j + \frac{1}{n}\sum\Delta v_i\Delta v_j
\label{eq:apA12}
\end{equation}
where
\begin{equation}
\frac{1}{n} \sum{\Delta v_i (v_j -\bar{v}_j)} = -\frac{2GM}{r_s^2w^3} \bigg[w^2C(x_i,v_j) - \sum_{k=x,y,z} w_iw_kC(x_k,v_j)\bigg].\label{eq:apA2}
\end{equation}
The last term in Eq.~\eqref{eq:apA12} is 
\begin{align}\label{eq:apA3}
\frac{1}{n} \sum{\Delta v_i \Delta  v_j} = \bigg(\frac{2GM}{r_s^2w^3}\bigg)^2\bigg[w^4C(x_i,x_j) & -  \sum_{k=x,y,z}\bigg( 
w^2w_k \big[w_j C(x_i,x_k) + w_i C(x_j,x_k)\big] -
w_iw_jw_k^2C(x_k, x_k)\nonumber
\bigg) \\
& +w_iw_j\bigg(2w_xw_yC(x,y) + 2w_xw_zC(x,z) + 2w_yw_zC(y,z) \bigg)\bigg].
\end{align}
These expressions take a much simpler form if the initial covariance matrix $\Sigma_{\varpi,0}$ (e.g. Eq.~\ref{eq:covariance}) is diagonal. In this case, Eq.~\eqref{eq:apA12} simplifies to
\begin{equation}
\frac{1}{n} \sum{\Delta v_i (x_j -\bar{x}_j)} = -\epsilon \frac{w}{r_s} C(x_j,x_j) \bigg[\delta_{ij} - \frac{w_iw_j}{w^2}\bigg],
\label{eqapp:delta_xivj}
\end{equation}
where 
\begin{equation}
    \epsilon = \frac{2GM}{r_s w^2}
\end{equation} 
is a unit less parameter and $\delta_{ij}$ is the Kronecker delta. Equation \eqref{eq:apA2} simply vanishes, since it only features off-diagonal terms, and Eq.~\eqref{eq:apA3} reduces to
\begin{equation}
\frac{1}{n} \sum{\Delta v_i \Delta v_j} = \epsilon^2 w^2 \bigg[\frac{C(x_i,x_j)}{r_s^2} -
\frac{w_iw_j}{w^2 r_s^2} \big[C(x_i,x_i) + C(x_j,x_j)\big]  +
\sum_{k=x,y,z}\bigg(\frac{w_iw_jw_k^2}{w^4}\frac{C(x_k, x_k)}{r_s^2}
\bigg) \bigg].
\label{eqapp:delta_vivj}
\end{equation}

In Appendix~\ref{Ap:expl_dep} we will turn back to these simplified expressions. It will be convenient to express Eqs.~\eqref{eqapp:delta_xivj} and \eqref{eqapp:delta_vivj} in terms of $\epsilon$ and another parameter (respectively $\Delta_{ij}$ and $D_{ij}$) that carries all other terms, such that in this specific case we can write
\begin{equation}
    \frac{1}{n} \sum{\Delta v_i (x_j -\bar{x}_j)} = -\epsilon \Delta_{ij},
    \label{eqapp:Deltaij}
\end{equation}
and
\begin{equation}
    \frac{1}{n} \sum{\Delta v_i \Delta v_j} = \epsilon^2 D_{ij}.
     \label{eqapp:Dij}
\end{equation}

\section{Subhalo scaling relations}\label{sec:subscarel}

For the scaling of the subhalos scale radius $r_s$ with mass $M$ we have used several scaling relations, they are listed below. To obtain the scaling we first relate the subhalo mass $M$ with the maximum circular velocity $V_{\rm max} = V_c(r_{\rm max})$.
In \cite{Springel2008} (see their Fig.~27) we find
\begin{equation}
    V_{\rm max} = \bigg(\frac{M}{3.37\cdot 10^7~{\rm M}_\odot}\bigg)^{1/3.49} \cdot 10~{\rm km/s},
\end{equation}
which is an empirical scaling relation based on the subhalos down to the mass-range of $\lesssim10^5~{\rm M}_\odot$, identified in the `Aquarius' simulations. 
Next, $r_{\rm max}$ is related to $V_{\rm max}$ using Eqs.~(6,8,9) from \citeauthor{Springel2008}
\begin{equation}
    r_{\rm max} = {V_{\rm max}}~\bigg[\frac{\delta_c H_0^2}{14.426} \bigg]^{-\frac{1}{2}}\cdot  0.62, \label{eq:rmaxvmax}
\end{equation}
where the factor $0.62$ is added based on the comment in the caption of Fig.~26 of \citet[][see also below]{Springel2008} and we assume $H_0 = 73~{\rm km/s/Mpc}$. The final missing piece is $\delta_c$, which is related to the concentration parameter $c$ by
\begin{equation}
    \delta_c = \frac{200}{3} c^3 ~ \bigg(\log(1+c)-\frac{c}{1+c}\bigg)^{-1}.
\end{equation}
Typically $c$ is related to the subhalo mass $M$, motivated by \citet{Springel2008}, we relate the two using an empirical scaling relation found by \cite{Neto2008TheConcentrations} for relaxed halos
\begin{equation}
    c = 5.26~\bigg(\frac{M}{h \cdot 10^{14}}\bigg)^{-0.10},
\end{equation}
where $h=H_0 / (100~{\rm km/s})$. We note that \citet{Springel2008} find that the resulting scaling relation of $V_{\rm max} \sim r_{\rm max}$ is lower than the relation found from extrapolating the results of \citeauthor{Neto2008TheConcentrations} (which is not calibrated for subhalos in the low-mass range that we consider here). The offset is $0.62$, which is why we add this factor~in~Eq.~\eqref{eq:rmaxvmax}. 

With the equations above we can relate the subhalo mass $M$ to $V_{\rm max}$ and a corresponding $r_{\rm max}$. The scale radius $r_{s,{\rm NFW}}$ is related to $r_{\rm max}$ simply as
\begin{align}
    r_{\rm max} = 2.163 \cdot r_{s,{\rm NFW}},
\end{align}
which is found numerically from calculating where $V_c(r_{\rm max}) = V_{\rm max}$ (but see Eq.~(11) of \cite{Diemand2007FormationSubstructure}, where we originally found the relation).

Finally, in this main text we use a Plummer profile to describe the subhalos, rather than an NFW. Therefore, we relate the scale radii of the two profiles by equation the acceleration at $r_{\rm max}$
\begin{equation}
    a_{\rm NFW}(r_{\rm max}) = a_{\rm Plummer}(r_{\rm max}) = -\frac{GM}{(r_{\rm max}^2 + r_s^2)^{3/2}}\cdot r_{\rm max}.
\end{equation}
The scale radius of the Plummer, $r_s$, can be found by solving the equation above, which then is a function of $M$ only. Finally, by fitting the scaling relation numerically we find that the scale radius depends on mass by $r_s \propto M^{0.397}\sim M^{2/5}$.

\section{Computation of the density contrast at late times}\label{Ap:expl_dep}

As described in the main paper, the density contrast at late times takes the form (see Eq.~\ref{eq:dens_plateau})
\begin{equation}
 \delta\rho^\mathrm{gap}_\mathrm{str} = \sqrt{\frac{\det{|\sigma_{\Theta_0}|}_\mathrm{str}}{\det{|\sigma_{\Theta_0}|}_\mathrm{gap}}}.
\label{eqapp:dens_plateau}
\end{equation}
To be able to establish its
dependence on the characteristic parameters of the encounter, we need to determine the form of determinant of the matrix
$\sigma_{\Theta_0}$. This is the upper left, 3$\times$3, submatrix of $\sigma_{\omega,0}$ that is described in 
Sec.~\ref{sec:centraldensity}. This latter matrix, following the notation of Sec.~\ref{sec:AA2orb}, takes the form
\begin{equation}
\sigma_{\omega,0} = {{\cal M}^{-1}_0}^\dagger\sigma_{\varpi,0}{\cal M}^{-1}_0,
\label{eqapp:sigma}
\end{equation}
where ${\cal M}_0$ is given by Eq.~\eqref{eq:tmatrix}, namely
$ {\cal M}_0 = {\cal M}^{\mathrm{AA} \leftarrow \mathrm{st}}_0 {\cal
  M}^{\mathrm{st} \leftarrow \mathrm{cyl}}_0 {\cal M}^{\mathrm{cyl}
  \leftarrow \mathrm{xyz}}_0$, and thus represents the coordinate
transformations from Cartesian to action-angle variables. For example
the matrix that accounts for the transformation from St\"ackel
coordinates to action angles,
${\cal M}^{\mathrm{AA} \leftarrow \mathrm{st}}_0$, contains the
derivatives of the characteristic function and its general form is
given in Eq.~(A2) of HW99. In Eq.~\eqref{eqapp:sigma}, $\sigma_\varpi^{-1} = \Sigma_\varpi$ is
the inverse of the 6$\times$6 covariance~matrix in Cartesian
coordinates. Therefore, $\sigma_{\Theta_0}$ for the stream depends on
location as well as on the initial properties of the stream, and
similarly for the gap.

The matrix $\sigma_{\varpi,0}$ is of the form
\begin{equation}
\sigma_{\varpi,0} =
\begin{pmatrix}
  \sigma_{\boldsymbol{x},0} & \sigma_{\boldsymbol{xv},0}\\
  \sigma^\dagger_{\boldsymbol{xv},0} & \sigma_{\boldsymbol{v},0}
 \end{pmatrix}, 
\label{eq:sigma0}
\end{equation}
and since we may express as
\begin{equation}
{\cal M}_0^{-1} = 
\begin{pmatrix}
  A & B\\
  C& D
 \end{pmatrix}, 
\end{equation}
this means that 
\begin{equation}
\sigma_{\Theta_0} = A^\dagger  \sigma_{\boldsymbol{x},0} A + C^\dagger \sigma^\dagger_{\boldsymbol{xv},0} A + A^\dagger  \sigma_{\boldsymbol{xv},0} C + C^\dagger \sigma^\dagger_{\boldsymbol{v},0} C.
\label{eq:theta_0}
\end{equation}
Using matrix inversions, if 
\begin{equation}
{\cal M}_0 = \begin{pmatrix}
  t_1 & t_2\\
  t_3 & t_4
 \end{pmatrix}, 
\end{equation}
and $t_4$ is invertible then
\begin{equation}
 A = (t_1 - t_2 t_4^{-1} t_3)^{-1},  \qquad {\rm and}  \qquad
C = -t_4^{-1} t_3 A = T_{43} A,
\label{eqapp:t43}
\end{equation} 
where recall that $A$ and $T_{43}$ represent coordinate
transformations, and therefore depend only on
location.  These matrices have been set to be identical for the gap
and the stream in Eq.~\eqref{eqapp:dens_plateau}, which significantly
simplifies subsequent computations. If we now replace in Eq.~\eqref{eq:theta_0}, this results in 
\begin{equation}
\det |\sigma_{\Theta_0}| = (\det A)^2 \det (\sigma_{\boldsymbol{x},0} + T_{43}^\dagger \sigma^\dagger_{\boldsymbol{xv},0} + \sigma_{\boldsymbol{xv},0} T_{43} + T_{43}^\dagger \sigma^\dagger_{\boldsymbol{v},0} T_{43}),
\label{eqapp:detTheta0}
\end{equation}
The expression for $\det |\sigma_{\Theta_0}|_{\rm str}$ using the
above equation has been worked out in detail by HW99 for a stream
generated from an initially isotropic Gaussian distribution in
configuration and velocity space, and for a preferred location along the orbit, namely
the apocentre. The reader may wish to check the explicit expressions
in the case of an axisymmetric system (the last equation in their
Appendix B), and for a system described with St\"ackel coordinates
(Eq.~C13 in their Appendix C).

We now proceed to determine the form of the submatrices of
$\sigma_{\varpi,0}$ given in Eq.~\eqref{eq:sigma0} and needed in
Eq.~\eqref{eqapp:detTheta0}. For the stream we will assume no initial
correlations between positions and velocities in the stream (i.e. a
diagonal covariance matrix $\Sigma_{\varpi,0}$), which means that the
submatrix $\sigma_{\boldsymbol{xv},0}^{\rm str} = 0$, and that
$\sigma_{{\boldsymbol x},0}^{\rm str} =
[\mathbb{C}_{{\boldsymbol{x},\boldsymbol{x}}_0}^{\rm str}]^{-1}$ and
$\sigma_{{\boldsymbol v},0}^{\rm str} =
[\mathbb{C}_{{\boldsymbol{v},\boldsymbol{v}}_0}^{\rm str}]^{-1}$
according to Eq.~\eqref{eq:covariance}. For the gap, we may express
\begin{equation}
\mathbb{C}_{{\boldsymbol{v},\boldsymbol{v}}_0}^{\rm gap} = \mathbb{C}_{{\boldsymbol{v},\boldsymbol{v}}_0}^{\rm str} + \epsilon^2 
\, \mathbf{D}
\label{eqapp:cvv}
\end{equation}
where the elements of $\mathbf{D}$ are given by $D_{ij}$, see Eq.~\eqref{eqapp:Dij}.
Furthermore, 
\begin{equation}
\mathbb{C}_{{\boldsymbol{x},\boldsymbol{v}}_0}^{\rm gap} = - \epsilon 
\, \mathbf{\Delta}
\label{eqapp:cxv}
\end{equation}
where the elements of $\mathbf{\Delta}$ are given by $\Delta_{ij}$, see Eq.~\eqref{eqapp:Deltaij}.

To compute the submatrices $\sigma_{{\boldsymbol x},0}^{\rm gap}$,
$\sigma_{\boldsymbol{xv},0}^{\rm gap}$ and
$\sigma_{{\boldsymbol v},0}^{\rm gap}$, in Eq.~\eqref{eqapp:detTheta0} we use that
$ {\sigma}_{\varpi,0}^{\rm gap} = {\Sigma_{\varpi,0}^{\rm gap}}^{-1}$, which is
given by Eq.~\eqref{eq:bigSigma_0}. The inverse of this block matrix can
be computed explicitly\footnote{see {\tt https://en.wikipedia.org/wiki/Block\_matrix\#Block\_matrix\_inversion} and references therein}, provided the matrix
$W = \mathbb{C}_{{\boldsymbol{v},\boldsymbol{v}}_0}^{\rm gap} -
{\mathbb{C}_{{\boldsymbol{x},\boldsymbol{v}}_0}^{\rm gap}}^{\dagger}
[\mathbb{C}_{{\boldsymbol{x},\boldsymbol{x}}_0}^{\rm gap}]^{-1}
\mathbb{C}_{{\boldsymbol{x},\boldsymbol{v}}_0}^{\rm gap}$ is
invertible. Using Eqs.~\eqref{eqapp:cvv} and \eqref{eqapp:cxv}, and to the
lowest order in $\epsilon$ we find that
\begin{eqnarray}
\sigma_{{\boldsymbol v},0}^{\rm gap} & =  & W^{-1} \approx [\mathbb{C}_{{\boldsymbol{v},\boldsymbol{v}}_0}^{\rm
  str}]^{-1} \bigg[\mathbf{I} - \epsilon^2 \,
(\mathbf{D} - \mathbf{\Delta}^\dagger \,
[\mathbb{C}_{{\boldsymbol{x},\boldsymbol{x}}_0}^{\rm str}]^{-1}
\mathbf{\Delta})[\mathbb{C}_{{\boldsymbol{v},\boldsymbol{v}}_0}^{\rm str}]^{-1}\bigg], \\
\sigma_{\boldsymbol{xv},0}^{\rm gap} & =  & \epsilon
\, [\mathbb{C}_{{\boldsymbol{x},\boldsymbol{x}}_0}^{\rm str}]^{-1} \mathbf{\Delta} \, [\mathbb{C}_{{\boldsymbol{v},\boldsymbol{v}}_0}^{\rm str}]^{-1} 
\\
\sigma_{{\boldsymbol x},0}^{\rm gap} & = &  [\mathbb{C}_{{\boldsymbol{x},\boldsymbol{x}}_0}^{\rm str}]^{-1}
+ \epsilon^2 \, [\mathbb{C}_{{\boldsymbol{x},\boldsymbol{x}}_0}^{\rm str}]^{-1} \mathbf{\Delta} \,
[\mathbb{C}_{{\boldsymbol{v},\boldsymbol{v}}_0}^{\rm str}]^{-1} \mathbf{\Delta}^\dagger  [\mathbb{C}_{{\boldsymbol{x},\boldsymbol{x}}_0}^{\rm str}]^{-1} .
\end{eqnarray}

We are now almost ready to compute the density contrast, since 
\begin{equation}
\det |\sigma_{\Theta_0}|_{\rm gap} \sim (\det A)^2 \, \det [\sigma_{{\boldsymbol x},0}^{\rm str} + 
T_{43}^\dagger \sigma_{{\boldsymbol v},0}^{\rm str} T_{43} + 
\epsilon \, (T_{43}^\dagger \sigma_{{\boldsymbol v},0}^{\rm str} \mathbf{\Delta}^\dagger \sigma_{{\boldsymbol x},0}^{\rm str} + \sigma_{{\boldsymbol x},0}^{\rm str} \mathbf{\Delta} \sigma_{{\boldsymbol v},0}^{\rm str} T_{43}) ] \sim \det  |\sigma_{\Theta_0}|_{\rm str} \, \det [\mathbf{I} + \epsilon \, \mathbf{K}],
\end{equation}
where the matrix $\mathbf{K}$ is 
\begin{equation}
\mathbf{K} = [T_{43}^\dagger \sigma_{{\boldsymbol v},0}^{\rm str} \, \mathbf{\Delta}^\dagger \sigma_{{\boldsymbol x},0}^{\rm str} + \sigma_{{\boldsymbol x},0}^{\rm str} \mathbf{\Delta} \, \sigma_{{\boldsymbol v},0}^{\rm str} T_{43} ] 
[\sigma_{{\boldsymbol x},0}^{\rm str} + 
T_{43}^\dagger \sigma_{{\boldsymbol v},0}^{\rm str} T_{43}]^{-1},
\end{equation}
and it is therefore dependent on the initial properties of the stream
(at the time of the encounter and through the $\sigma$ matrices), the
location of the encounter in phase-space (through the matrix $T_{43}$
introduced in Eq.~\ref{eqapp:t43}), and the characteristic parameters
of the encounter (through the matrix ${\bf \Delta}$ whose elements can be described as $\frac{w}{r_s}$ times some function $g(\mathbb{C}_{{\boldsymbol{x},\boldsymbol{x}}_0}^{\rm str},\alpha,\theta)$, see Eqs.~\ref{eqapp:delta_xivj} and \ref{eqapp:Deltaij}).
Since we may
express
$\det [\mathbf{I} + \epsilon \, \mathbf{K}] \sim 1 + \epsilon \, {\rm
  tr} \mathbf{K}$, the density contrast at late times becomes 
\begin{equation}
\delta\rho^\mathrm{gap}_\mathrm{str} 
\sim \sqrt{\frac{\det{|\sigma_{\Theta_0}|}_\mathrm{str}}{\det{|\sigma_{\Theta_0}|}_\mathrm{gap}}} 
\sim (1 +  \epsilon \, {\rm tr} \, \mathbf{K})^{-1/2} 
\sim 1 - \frac{\epsilon}{2}  {\rm tr} \, \mathbf{K}
\sim 1 - \frac{GM}{r_s^2 w} f(\theta, \alpha, \mathbb{C}^{\rm
  str}_{{\boldsymbol{x},\boldsymbol{x}}_0}, \mathbb{C}^{\rm
  str}_{{\boldsymbol{v},\boldsymbol{v}}_0},
\boldsymbol{x}_0,\boldsymbol{v}_0),
\label{eqapp_rho}
\end{equation}
where $f(\theta, \alpha, \mathbb{C}^{\rm
  str}_{{\boldsymbol{x},\boldsymbol{x}}_0}, \mathbb{C}^{\rm
  str}_{{\boldsymbol{v},\boldsymbol{v}}_0},
\boldsymbol{x}_0,\boldsymbol{v}_0)$ is a function that depends on the impact configuration and initial state of the stream.

\end{document}